\documentclass[pra, aps, 10pt, twocolumn, floatfix,superscriptaddress]{revtex4-1} %%

\usepackage {graphicx}
\usepackage {amssymb}
\usepackage {amsmath}
\usepackage {verbatim}
\usepackage{color}
\usepackage[normalem]{ulem}

\begin{document}

\title{Area theorem in a ring laser cavity}

\author{Anton Pakhomov}

\affiliation{St. Petersburg State University, Universitetskaya nab. 7/9, St. Petersburg 199034, Russia}

\author{Mikhail Arkhipov}

\affiliation{St. Petersburg State University, Universitetskaya nab. 7/9, St. Petersburg 199034, Russia}

\affiliation{Ioffe Institute, Politekhnicheskaya str. 26, St. Petersburg 194021, Russia}

\author{Nikolay Rosanov}

\affiliation{St. Petersburg State University, Universitetskaya nab. 7/9, St. Petersburg 199034, Russia}

\affiliation{Ioffe Institute, Politekhnicheskaya str. 26, St. Petersburg 194021, Russia}

\author{Rostislav Arkhipov}

\affiliation{St. Petersburg State University, Universitetskaya nab. 7/9, St. Petersburg 199034, Russia}

\affiliation{Ioffe Institute, Politekhnicheskaya str. 26, St. Petersburg 194021, Russia}

\begin{abstract}
The generalization of the area theorem is derived for the case of a pulse circulating inside a ring laser cavity. In contrast to the standard area theorem, which is valid for a single pass of a traveling pulse through a resonant medium, the obtained generalized area theorem takes into account the medium-assisted nonlinear self-action effects through the medium excitation left by the pulse at the previous round-trip in the cavity. The generalized area theorem was then applied to the theoretical description of the dynamics of a single-section ring-cavity laser and the steady solutions for the pulse area and for the medium parameters were found both in the limit of a lumped model and for a spatially-extended system. The derived area theorem can be used for the convenient analytical description of different coherent photonic devices, like coherently mode-locked lasers or pulse compressors, as well as for the analysis of the photon echo formation in cavity-based setups.
\end{abstract}

\maketitle

\section{Introduction}

Over last decades there has been an impressive progress achieved in the generation of ultra-short pulses of femtosecond and even attosecond duration \cite{Krausz, Chini, Mourou_Nobel, Xue_2022, Midorikawa}. So short pulses allow to explore many unusual regimes of the light-matter interaction as well as unveil the dynamics of different ultrafast processes in atoms, molecules and solids \cite{Biegert, HeSun, Hui}.

When the pulse duration is much smaller than the coherence / polarization relaxation time $T_2$ in the medium (also called the dephasing time), the resulting light-matter interaction is called the coherent one \cite{Allen, kryukov1970propagation, poluektov1975self, gl1971analytical}. 
Typically the dephasing time $T_2$ in gaseous media falls into the ns range, while in solids $T_2$ at the room temperature varies from several tens to several hundreds of fs. This means that few- and subcycle pulses in the optical range, which are routinely generated nowadays, would surely interact coherently with most resonant media.

Over such coherent light-matter interaction the phase matching between multiple resonant centers in the medium induced by the driving field is preserved, what gives rise to a number of fascinating optical phenomena. As the most prominent seems the self-induced transparency (SIT) phenomena, when a light pulse propagates through an absorbing resonant medium without losses and with its shape unchanged \cite{McCall, McCall_2}.

Since its discovery, SIT phenomena was experimentally observed in many different media \cite{poluektov1975self, borri2002rabi, Beham, Choi, Marini, karni2013rabi, kolarczik2013quantum, Nandi}.
%Studies of SIT in the few- and single-cycle regimes have led to the prediction of novel phenomena of the light-matter interaction, such as half-cycle SIT soliton existence in two-level medium \cite{Bullough,Kalosha}, ultra-short SIT plasmon solitons in a metal film \cite{Marini}, the light self-stopping in a homogeneous medium \cite{arkhipov2022self}, light-induced dynamical microcavities formation \cite{diachkova2023light} etc. 
Still the practical applications of SIT have been quite poor (see the review \cite{rosanov2021dissipative} and references therein). For instance, several studies have demonstrated the possibility of the compression of ultra-short pulses based on the SIT phenomena \cite{arkhipov2021single}.  However, in the last years the so-called  SIT- or coherent laser mode-locking (CML) is actively studied \cite{Kozlov_CML, Menyuk, Talukder, Kozlov_CML_2, Kozlov_CML_3,Arkhipov_JETPL_2015,arkhipov2015modeling,Arkhipov_OptCommun, Arkhipov_OL_2016, Arkhipov_SciRep_2021, Arkhipov_PRA_2022, Outafat, Pakhomov_PRA_2023, ArkhipovMV_JETPL_2015,ArkhipovMV_LPL_2018,Arkhipov_JETPL_2019,ArkhipovMV_JPCS,ArkhipovMV_PRA_2020}, which looks as the most promising application of the SIT phenomena. CML represents an alternative mechanism of the passive mode-locking in a laser cavity based on the formation of SIT solitons. Due to the coherent pulse-medium interaction in such mode-locked lasers, some significant improvements can be achieved as compared to standard passively mode-locked lasers with saturable absorbers.

Namely, standard passively mode-locked lasers rely on the incoherent saturation of the absorber, leading to the generated pulses  being limited in duration by the value of the coherence relaxation time $T_2$ in the active medium \cite{Keller_Nature, Keller_Nature, Rafailov, Keller, Diels}. At the same time the SIT mode-locking does not exhibit such restrictions. Instead, since the generated pulses coherently interact with the active medium, the coherent mode-locking allows to produce pulses much shorter in duration than the coherence relaxation time $T_2$ \cite{Kozlov_CML, Menyuk, Talukder, Kozlov_CML_2, Kozlov_CML_3,Arkhipov_JETPL_2015,arkhipov2015modeling,Arkhipov_OptCommun, Arkhipov_OL_2016, Arkhipov_SciRep_2021, Arkhipov_PRA_2022, Outafat, Pakhomov_PRA_2023, ArkhipovMV_JETPL_2015,ArkhipovMV_LPL_2018,Arkhipov_JETPL_2019,ArkhipovMV_JPCS,ArkhipovMV_PRA_2020}.
Moreover, in some  configurations of coherently mode-locked lasers the pulses up to a single-cycle duration were obtained \cite{Kozlov_CML_2, Kozlov_CML_3, Arkhipov_PRA_2022}. It is also worth noting that the CML regime was recently experimentally demonstrated, though with the coherent absorber section only \cite{ArkhipovMV_JETPL_2015,ArkhipovMV_LPL_2018,Arkhipov_JETPL_2019,ArkhipovMV_JPCS,ArkhipovMV_PRA_2020}.

The pulse area theorem represents a unique analytical tool to describe the coherent pulse propagation in resonant media. For the first time the area theorem was derived in the pioneering work by McCall and Hahn \cite{McCall} for a single pass of a pulse through a two-level medium. The area theorem holds for ultra-short pulses up to few-cycle ones, but breaks down for the propagation of large-area few-cycle and especially subcycle pulses in a two-level medium \cite{Hughes,  Song_Gong, Novitsky, arkhipov2021coherent}.

Some generalizations of the standard area theorem have been developed in recent years, mainly for the description of the photon echo. Namely, the extensions of the area theorem were derived for three-level atoms \cite{Eberly_Kozlov, Shchedrin}, two-pulse photon echo \cite{Urmancheev}, tripodal four-level atomic system interacting with three fields in resonance \cite{Gutierrez} and optically dense media \cite{Moiseev_PRR}. Specifically for cavity setups, the area theorem was considered for a single-mode ring cavity \cite{Chaneliere, Moiseev_ring} and a single-mode Fabry-Perot cavity \cite{Moiseev_OL}. However, in these papers only one longitudinal cavity mode was assumed to be excited and thus no propagation effects were considered. The area theorem with the propagation included was studied in a single-mode waveguide \cite{Moiseev_PRA_2023}, but only for a single pass of  propagating pulses.

At the same time, the proper theoretical description of coherently mode-locked lasers or pulse compressors with cavity setups requires the revision of the standard area theorem. Indeed, in a laser cavity a traveling pulse passes through intracavity media at each round-trip, so that the interaction dynamics each time depends on the outcome of the previous round-trip. Specifically, in coherently mode-locked lasers the medium relaxation times are often longer than the cavity round-trip time. As the result, the medium excitation left after the pulse passage affects the pulse propagation at the next round-trip. That is why the respective extension of the area theorem is needed.
The standard area theorem was earlier applied to the analysis of the CML regime in lasers in Refs. \cite{arkhipov2015modeling, Arkhipov_OL_2016, Arkhipov_SciRep_2021, Pakhomov_PRA_2023}, but all the above described medium-mediated self-action effects were ignored.

It is worth mentioning that in the theoretical treatment of the standard passive mode-locking with saturable absorbers the medium polarization is always adiabatically eliminated  from the model equations 
\cite{haus1975theory, haus1975theory_2, haus2000mode, new1974pulse, kartner1996soliton, kurtner1998mode, paschotta2001passive, vladimirov2004new, vladimirov2004delay, vladimirov2005model}.
%\cite{haus1975theory, haus1975theory_2, haus2000mode, new1974pulse, kartner1996soliton, kurtner1998mode, paschotta2001passive, vladimirov2004new, vladimirov2004delay, vladimirov2005model, arkhipov2012hybrid, arkhipov2015pulse, arkhipov2016semiconductor, vladimirov2021delay,vladimirov2022short}. 
As the result, the coherent effects of the light-matter interaction are not taken into account. However, today's solid-state laser devices make it possible to generate pulses of femtosecond durations \cite{song2020recent, han2020generation, liu2021short, thompson2009ingaas, rafailov2005high, yadav2023edge}, what is comparable to the medium coherence lifetime $T_2$ in such lasers. Thus, in such a case conventional mode-locking theories are not applicable for the correct description of the mode-locking regime.  % Next,  it was shown in \cite{Arkhipov_SciRep_2021, Arkhipov_PRA_2022} that to obtain shorter pulses, cavity length must be reduced according to scaling rules. In paticular, it is possible to generate single-cycle pulses in laser with ultra-short cavity with an ultra-high repetition rate via CML regime \cite{Arkhipov_PRA_2022}.
%In modern compact laser systems coherent effects are of crucial importance since pulse duration is shorter than lasing medium coherence time $T_2$.   
For this reason, the practical implementation of such ultrafast laser sources requires modification of existing mode-locking theories by significantly new theoretical approaches, including the proper treatment of the coherent light-matter interaction and the pulse area evolution in the cavity.

In this paper, we derive in a closed analytical form the generalized area theorem for a pulse circulating inside a ring cavity. We specifically address several important cases differing by the relative values of the cavity round-trip time and the medium relaxation constants. The derived area theorem is then applied for the proper description of the dynamical regimes of a single-section ring-cavity laser.

The paper is organized as follows. In Section II we present the considered model and derive the governing equations for the evolution of a two-level medium during a single round-trip of a traveling pulse in a ring cavity. In Section III we formulate the generalized area theorem for the pulse propagation in a ring cavity and examine several approaches to reduce the derived area theorem including the fast coherence relaxation and the system with the lumped parameters. In Section IV we apply the generalized area theorem to a single-section laser setup with the gain medium in the cavity and analyse the steady-state solutions and their stability for a lumped model. Section V is devoted to the case of a spatially-extended gain medium and we derive explicit analytical expressions for the pulse area and the population inversion in the medium. Finally, paper summary and concluding remarks are given in Section VI.

\section{Model}

The most important quantity used for the description of the coherent pulse interaction with a two-level resonant medium is the so-called pulse area, given as \citep{Allen}:
\begin{equation}
\Phi(z)=\frac{d_{12}}{\hbar} \int_{-\infty}^{+\infty} \mathcal{E}(t',z) dt',
\label{eqPhi}
\end{equation}
where $\mathcal{E}(t,z)$ is the slowly varying envelope of the electric field in the pulse, $d_{12}$ is the transition dipole moment of the two-level medium, $\hbar$ is the reduced Planck constant and the time-domain integration is performed over the whole pulse duration.

If a propagating light pulse coherently interacts with an inhomogeneously broadened two-level medium, the pulse area evolution can be described by a simple differential equation, which is called the area theorem. The standard form of this theorem originally derived by McCall and Hahn is as follows \citep{McCall_2, kryukov1970propagation, Allen, Eberly_98}:
\begin{eqnarray}
\frac{d\Phi}{dz} = \pm \frac{4 \pi^2 N_0 d^2_{12} \omega_{12} g(0)}{n_{\text{ph}} \hbar c} \ \sin \Phi,
\label{AreaTheorem}
\end{eqnarray}
where the plus sign in the right-hand side corresponds to an amplifying medium, while the minus sign refers to an absorbing medium. Other quantities are the concentration of two-level resonant centers $N_0$, the medium transition frequency $\omega_{12}$, the host refractive index $n_{\text{ph}}$
and the function $g(\Delta)$ describes the properly rescaled inhomogeneously broadened line, so that:
\begin{equation*}
\nonumber
\int_{-\infty}^{+\infty} g(\Delta) \ d\Delta = 1, 
\label{inhomogen_line}
\end{equation*}
where $\Delta$ is the frequency detuning from the center of the inhomogeneously broadened line.

Eq.~\eqref{AreaTheorem} was obtained under the assumption that there was no induced medium polarization by the arrival of the propagating pulse and the medium was either fully inverted (i.e. all $N_0$ resonant centers per unit volume  were in the excited state) or, inversely, fully uninverted (all $N_0$ resonant centers per unit volume  were in the ground state). Eq.~\eqref{AreaTheorem} has the following general solution:
\begin{equation}
\tan \Big( \frac{\Phi}{2} \Big) = \tan \Big( \frac{\Phi_0}{2} \Big) \cdot e^{ \pm \alpha N_0 z},
\label{AreaTheoremSolution}
\end{equation}
where $\Phi_0$ is the initial pulse area and we have denoted:
\begin{equation}
\alpha = \frac{4 \pi^2 d^2_{12} \omega_{12} g(0)}{n_{\text{ph}} \hbar c}.
\label{alpha_def}
\end{equation}

This solution predicts the steady values of the pulse area equal to integer numbers of $\pi$. In fact, in an absorbing two-level medium the stable SIT-soliton must have the pulse area equal to 2$\pi$ \cite{McCall_2, Allen}, and is thus usually called the $2 \pi$-pulse. Such propagating $2 \pi$-pulse excites the medium at its leading edge and then returns back to the ground state at the pulse's trailing edge. The absorbing medium therefore undergoes a single Rabi flopping over the pulse duration. As the medium behind such a SIT-soliton stays fully uninverted, the $2 \pi$-pulse experiences almost no losses.

In contrast, in the amplifying / gain medium the pulse with $\Phi = 2 \pi$ appears to be unstable, while the only stable value of the pulse area equals $\Phi = \pi$. For any $\Phi_0 \in (0; 2\pi)$ the pulse approaches the value $\Phi = \pi$ upon its propagation in the medium. At the same time, the steady state $\Phi = \pi$ is unstable in the absorbing medium and  the propagating pulse either approaches the pulse area $\Phi = 0$ for $\Phi_0 \in (0; \pi)$ or approaches the value of the pulse area $\Phi = 2 \pi$ for $\Phi_0 \in (\pi; 2\pi)$.

In order to provide more general treatment of the coherent pulse propagation in a ring laser cavity, we are, however, interested to start off with the underlying equations for the system dynamics rather than the final result Eq.~\eqref{AreaTheorem}. We apply the system of Maxwell-Bloch equations for two-level gain and absorber media together with the usual wave equation for the electric field (magnetic properties of the media are neglected).

As we expect long multi-cycle mode-locked pulses, we make use of slowly-varying envelope (SVEA) and rotating-wave (RWA) approximations for the electric field. For simplicity, we also suppose linearly polarized electric field and neglect the transverse effects / diffraction. Thereby the problem reduces to the scalar equation for the slowly varying envelope of the electric field \cite{Allen, kryukov1970propagation}:
\begin{eqnarray}
\nonumber
\frac{\partial E}{\partial t} + \frac{c}{n_{\text{gr}}} \ \frac{\partial E}{\partial z} &=&  \frac{2 \pi \omega_{12}}{n_{\text{gr}} n_{\text{ph}}}\int_{-\infty}^{+\infty} g(\Delta) \ P_{a/g}(\Delta, t, z) \ d\Delta, \\
\label{WaveEq_SVEA}
\end{eqnarray} 
where $E$ is the slowly-varying envelope of the electric field, $P_{a/g}$ are the slowly-varying envelopes of the polarization of the absorber/gain media phase-shifted by $\pi / 2$ with respect to the electric field envelope, $n_{\text{gr}}$ and $n_{\text{ph}}$ are the group and the phase refractive indices of the host medium at the pulse carrier frequency $\omega_{12}$.

Let us multiply both sides of Eq.~\eqref{WaveEq_SVEA} by the factor $d_{12}/\hbar$ and integrate both sides over the whole pulse duration. As the electric field turns to zero in front of and beyond the propagating pulse, the first term in the left-hand side of Eq.~\eqref{WaveEq_SVEA} vanishes. As the result, Eq.~\eqref{WaveEq_SVEA} turns into:
\begin{eqnarray}
\nonumber
\frac{\partial \Phi}{\partial z} &=&  \frac{2 \pi \omega_{12} d_{12}}{n_{\text{ph}} \hbar c } \int_{-\infty}^{+\infty} g(\Delta) \ d\Delta  \ \int_{-\infty}^{+\infty} P_{a/g}(\Delta, t', z) dt', \\
\label{WaveEq_SVEA_Phi}
\end{eqnarray}
where the time-domain integration is performed over the whole pulse duration.

The key assumption made upon the derivation of the area theorem is related to the values of the pulse duration $\tau$, the coherence relaxation time $T_2$ (equal to the inversed width of the homogeneously broadened line) and the inversed width of the inhomogeneously broadened line $\tilde \Delta$. Namely, the following inequalities have to  hold:
\begin{eqnarray}
\tilde \Delta^{-1}  \ll  \tau  \ll  T_2.
\label{area_theorem_assume}
\end{eqnarray}
It is should be noted that due to the strong inhomogeneous line broadening as given by Eq.~\eqref{area_theorem_assume} 
the residual macroscopic medium polarization in the right-hand side of Eq.~\eqref{WaveEq_SVEA} 
left after the pulse passage vanishes over a time interval of the order of $\tilde \Delta^{-1}$ due to the dephasing between multiple oscillating dipoles across the inhomogeneously  broadened line (the so-called free polarization decay) \cite{Allen}. According to Eq.~\eqref{area_theorem_assume}, this means that, as the traveling pulse is gone, the residual macroscopic polarization in the right-hand side of Eq.~\eqref{WaveEq_SVEA} disappears much faster than the pulse extends, so that no other field is to be emitted afterwards.

With the conditions Eq.~\eqref{area_theorem_assume}, as it was shown in \cite{McCall_2, Allen}, the integral in the right-hand side of Eq.~\eqref{WaveEq_SVEA_Phi} can be simplified as:
\begin{eqnarray}
\nonumber
 \int_{-\infty}^{+\infty} g(\Delta) \ d\Delta   \int_{-\infty}^{+\infty} P_{a/g}(\Delta, t', z) \ dt' \\
 \to  2 \pi \ g(0) \ P_{a/g}(0, t_0 + \tau, z),
\label{WaveEq_SVEA_Phi_integral}
\end{eqnarray}
where $t_0$ is the time point, when the traveling pulse arrives to the spatial point $z$. The last factor in the right-hand side of Eq.~\eqref{WaveEq_SVEA_Phi_integral} 
thus represents the induced medium polarization at the central frequency of the medium's inhomogeneously broadened line (i.e. zero frequency detuning $\Delta = 0$) at the spatial point $z$ just after the pulse has fully passed this point. Eq.~\eqref{WaveEq_SVEA_Phi} for the pulse area evolution using the expression Eq.~\eqref{WaveEq_SVEA_Phi_integral} becomes:
\begin{eqnarray}
\frac{d \Phi}{d z} &=&  \frac{\alpha}{d_{12}} \ P_{a/g}(0, t_0 + \tau, z),
\label{WaveEq_SVEA_Phi_final}
\end{eqnarray}
with the parameter $\alpha$ defined by Eq.~\eqref{alpha_def}. Finally, the induced medium polarization in the right-hand side of Eq.~\eqref{WaveEq_SVEA_Phi_final} has to be found through the solution of the Bloch equations for the dynamics of the resonant medium.

The Bloch equations for a two-level resonant medium for the zero frequency detuning $\Delta = 0$ read as follows \cite{Allen}:
\begin{eqnarray}
\nonumber
\frac{d P_{a/g}}{d t} + \frac{P_{a/g}}{T_{2}} &=&  \frac{ d_{12}^2}{\hbar} \ N_{a/g} E, \\
\frac{d N_{a/g}}{d t} + \frac{N_{a/g} - N_{0, a/g}}{T_{1}} &=& - \frac{1}{ \hbar} \ P_{a/g} E,
\label{Bloch_0}
\end{eqnarray} 
where $N_{a/g}$ are the population inversions in the absorber/gain sections, $N_{0, a/g}$ are the equilibrium values of the population inversion, i.e. the pumping rates in absorber/gain sections, $T_{1}$ is the lifetime of the excited state, $T_{2}$ is the medium coherence lifetime.

It should be noted that the two-level model in Eq.~\eqref{Bloch_0} was successfully applied for the theoretical description of the coherent pulse propagation and the self-induced transparency phenomena in different gaseous and solid-state media \citep{Allen, kryukov1970propagation}. Moreover, this model was even able to describe the experimentally observed features of the Rabi flopping in such complex solid-state media, as bulk semiconductors \cite{Mucke, Mucke_2, Wegener}.

The coherent pulse propagation requires the pulse duration $\tau$ to be much smaller than the relaxation times $T_1, T_2$ in the medium. In this case we can reliably neglect the relaxation terms in the left-hand side of the Bloch equations Eqs.~\eqref{Bloch_0} upon the medium interaction with the pulse. The solution of  Eqs.~\eqref{Bloch_0} without relaxation can be then readily found as:
\begin{eqnarray}
\nonumber
N_{a/g} (t) &=& N_0 \ \cos \Big( \frac{ d_{12}}{\hbar} \int_{-\infty}^t E(t') dt'  +  \Theta_0 \Big),  \\
\nonumber
P_{a/g} (t) &=& d_{12} N_0 \ \sin \Big( \frac{ d_{12} }{\hbar} \int_{-\infty}^t E(t') dt'  +  \Theta_0 \Big),  \\
\label{Bloch_1}
\end{eqnarray}
with the spatial density $N_0$ and the integration constant $\Theta_0$ determined by the initial conditions.

For example, for an initially unexcited medium, i.e. the medium is in the ground state before the arrival of a driving pulse (absorber), one needs:
\begin{eqnarray}
\nonumber
N (t) \Big|_{t = -\infty} &=& - N_0 ,  \\
P (t) \Big|_{t = -\infty} &=& 0.
\label{Initial_1}
\end{eqnarray}
This case corresponds to the $\Theta_0 = \pi$ in Eq.~\eqref{Bloch_1}. In the opposite case of an initially fully excited medium, i.e. the medium is in the upper excited state before the arrival of a driving pulse (gain), one needs:
\begin{eqnarray}
\nonumber
N (t) \Big|_{t = -\infty}  &=&  N_0 ,  \\
P (t) \Big|_{t = -\infty}  &=&  0,
\label{Initial_2}
\end{eqnarray}
what already corresponds to the $\Theta_0 = 0$ in Eq.~\eqref{Bloch_1}. In these cases Eqs.~\eqref{Initial_1}-\eqref{Initial_2}, if inserting the respective solution for $P_{a/g} (t)$ from Eq.~\eqref{Bloch_1} to the spatial evolution equation Eq.~\eqref{WaveEq_SVEA_Phi_final}, one gets again the standard area theorem Eq.~\eqref{AreaTheorem}.

If the induced medium polarization does not fully relax over a full round-trip in the cavity, then the non-zero initial values of the polarization $P (t)$ arise, leading to other possible values of the integration constant $\Theta_0$ in Eq.~\eqref{Bloch_1}.
Specifically, for arbitrary initial conditions:
\begin{eqnarray}
\nonumber
N_{a/g} (t) \Big|_{t = -\infty}  &=&  N_{a/g}^{\text{init}},  \\
P_{a/g} (t) \Big|_{t = -\infty}  &=&  P_{a/g}^{\text{init}},
\label{Initial_3}
\end{eqnarray}
with the initial values of the quantities $N_{a/g}^{\text{init}}, \ P_{a/g}^{\text{init}}$,
the solution Eq.~\eqref{Bloch_1} turns to:
\begin{eqnarray}
\nonumber
N_{a/g} (t) &=& N_{a/g}^{\text{init}} \ \cos \Big( \frac{ d_{12}}{\hbar} \int_{-\infty}^t E(t') dt' \Big) - \\
\nonumber
&& \frac{P_{a/g}^{\text{init}}}{d_{12}} \sin \Big( \frac{ d_{12} }{\hbar} \int_{-\infty}^t E(t') dt' \Big),  \\
\nonumber
P_{a/g} (t) &=& P_{a/g}^{\text{init}} \cos \Big( \frac{ d_{12}}{\hbar} \int_{-\infty}^t E(t') dt' \Big)  +  \\
\nonumber
&& d_{12} N_{a/g}^{\text{init}} \ \sin \Big( \frac{ d_{12} }{\hbar} \int_{-\infty}^t E(t') dt' \Big).  \\
\label{Bloch_full}
\end{eqnarray}

Next, let us find out the solution of the Bloch equations Eqs.~\eqref{Bloch_0} in between the passages of a pulse circulating in a cavity. For this case we assume zero value of the electric field, i.e. skip the right-hand side in Eqs.~\eqref{Bloch_0}, and take into account the relaxation terms. The respective solutions of Eqs.~\eqref{Bloch_0} then yield:
\begin{eqnarray}
\nonumber
N_{a/g} (t) &=& N_{a/g}^{\text{init}} \ e^{- t / T_1} +  N_{0, a/g}  \Big( 1 - e^{- t / T_1} \Big), \\
\nonumber
P_{a/g} (t) &=& P_{a/g}^{\text{init}} \ e^{- t / T_2},  \\
\label{Bloch_2}
\end{eqnarray}
with the initial values $N_{a/g}^{\text{init}}, \ P_{a/g}^{\text{init}}$. Altogether Eqs.~\eqref{Bloch_full}-\eqref{Bloch_2} fully describe the medium evolution over the whole round-trip of a pulse inside a ring laser cavity.

\section{Generalized area theorem}

Having obtained the explicit expressions for the resonant medium evolution both during the pulse passage Eq.~\eqref{Bloch_full} and in between the sequential pulse passages Eq.~\eqref{Bloch_2}, we are now ready to formulate the generalized pulse area theorem for the unidirectional pulse propagation in a ring cavity.

Instead of the single  unknown $\Phi (z)$ as in the standard area theorem Eq.~\eqref{AreaTheorem}, in the general case we have to also consider the spatial distribution of the population inversion in the active laser medium $N (z)$ and the spatial distribution of the induced medium polarization $P(z)$.

Let us assume a ring cavity with the round-trip time $T_{\text{rt}}$. We suppose that a single isolated pulse circulates inside a cavity in one specific direction, e.g. anti-clockwise. Thus, the medium parameters at each spatial point $z$ evolve according to Eq.~\eqref{Bloch_full} as the pulse passes through this point, then decay according to 
Eq.~\eqref{Bloch_2} during the round-trip time 
$T_{\text{rt}}$ until the circulating pulse arrives again and so on.

We denote the pulse area at the $n$-th round-trip as $\Phi_n (z)$ and the population inversion and the induced medium polarization after $n$ full round-trips right before the pulse arrives to the point $z$ at the $(n+1)$-th round-trip as $N_n (z)$ and $P_n (z)$. Now bringing together Eqs.~\eqref{Bloch_full}-\eqref{Bloch_2} and the spatial evolution equation Eq.~\eqref{WaveEq_SVEA_Phi_final}, we get the following generalized pulse area theorem for the unidirectional pulse circulation inside a ring cavity:
\begin{eqnarray}
\nonumber
\frac{d \Phi_{n+1} (z)}{dz}  &=&  \alpha \Big( N_n (z) \sin \Phi_{n+1} (z) + \\
\nonumber
&&  \frac{P_n (z)}{d_{12}} \cos \Phi_{n+1} (z) \Big), \\
\nonumber
N_{n+1} (z)  &=&  \Big( N_n (z) \cos \Phi_{n+1} (z) - \\
\nonumber
&& \frac{P_n (z)}{d_{12}} \sin \Phi_{n+1} (z) \Big) \ e^{-T_{\text{rt}} / T_1 }  +  \\
\nonumber
&&  N_{0, g}  \Big( 1 - e^{-T_{\text{rt}} / T_1 } \Big), \\
\nonumber
P_{n+1} (z) &=& \Big( P_n (z) \cos \Phi_{n+1} (z) + \\
\nonumber
&&  d_{12} N_n (z) \sin \Phi_{n+1} (z) \Big)   e^{-T_{\text{rt}} / T_2 },  \\
\nonumber
0  &\le&  z  \le  L_{\text{cav}},  \\
\Phi_{n+1} (0)  &=&  \Phi_{n} (L_{\text{cav}}),
\label{pulse_area_gener}
\end{eqnarray}
with the cavity length $L_{\text{cav}}$.

One can easily see that in the limit $T_1, \ T_2  \ll  T_{\text{rt}}$  the population inversion and the induced medium polarization completely relax over the round-trip time to their stable values $N_n (z) = N_{0, g}$ and $P_n (z) = 0$ respectively. As the result, the generalized pulse area theorem Eq.~\eqref{pulse_area_gener} simply reduces to the classic one:
 \begin{eqnarray}
\nonumber
\frac{d \Phi_{n+1} (z)}{dz}  &=&  \alpha N_{0, g} \sin \Phi_{n+1} (z), \\
0  &\le&  z  \le  L_{\text{cav}}, 
\label{pulse_area_gener_classic}
\end{eqnarray}
i.e. exactly coincides with Eq.~\eqref{AreaTheorem}.

Typically, especially in solid media, the lifetime of the excited state $T_1$ is by many orders of magnitude greater than the coherence lifetime $T_2$. Therefore it is also of interest to consider the specific case with $T_2 \ll T_{\text{rt}} \lesssim T_1$. In this case the induced medium polarization fully vanishes over the round-trip time, so that we can put $P_n(z) = 0$ in Eq.~\eqref{pulse_area_gener}. As the result, the generalized area theorem gets reduced to the form:
\begin{eqnarray}
\nonumber
\frac{d \Phi_{n+1} (z)}{dz}  &=&  \alpha N_n (z) \sin \Phi_{n+1} (z), \\
\nonumber
N_{n+1} (z)  &=&  N_n (z) \cos \Phi_{n+1} (z)  \  e^{-T_{\text{rt}} / T_1 }  +  \\
\nonumber
&&  N_{0, g}  \Big( 1 - e^{-T_{\text{rt}} / T_1 } \Big), \\
\nonumber
0  &\le&  z  \le  L_{\text{cav}},  \\
\Phi_{n+1} (0)  &=&  \Phi_{n} (L_{\text{cav}}).
\label{pulse_area_gener_reduced}
\end{eqnarray}

The first equation for $\Phi_{n+1} (z)$ in Eq.~\eqref{pulse_area_gener_reduced} can be readily integrated and yields:
\begin{eqnarray}
\nonumber
\tan \Big(\frac{\Phi_{n+1} (z)}{2} \Big)  &=& \tan \Big(\frac{\Phi_{n+1} (0)}{2} \Big) \cdot e^{ \alpha \int_0^z N_n (z') dz'},  \\
\nonumber
0  &\le&  z  \le  L_{\text{cav}},  \\
\Phi_{n+1} (0)  &=&  \Phi_{n} (L_{\text{cav}}).
\label{pulse_area_gener_reduced_solution}
\end{eqnarray}
Here the function $N_n (z)$ under the integral sign is fully determined by the pulse area dependences in the previous round-trips $\Phi_{j} (z), \ j \le n.$ For instance, the function $N_n (z)$ according to the second equation in Eq.~\eqref{pulse_area_gener_reduced} can be expressed as:
\begin{eqnarray}
\nonumber
N_{n} (z)  &=&  N_{n-1} (z) \cos \Phi_{n} (z)  \  e^{-T_{\text{rt}} / T_1 }  +  \\
\nonumber
&&  N_{0, g}  \Big( 1 - e^{-T_{\text{rt}} / T_1 } \Big),
\label{pulse_area_gener_reduced_Nn}
\end{eqnarray}
where the function $N_{n-1} (z)$ is to be similarly expressed through $N_{n-2} (z)$ and $\Phi_{n-1} (z)$ and so on.

In the general case Eq.~\eqref{pulse_area_gener} the differential equation for the pulse area $\Phi_{n+1} (z)$ can be represented as:
\begin{eqnarray}
\nonumber
\frac{d \Phi_{n+1} (z)}{dz}  &=&  \alpha \  \sqrt{N^2_n (z) + \frac{P^2_n (z)}{d^2_{12}}} \sin \Big(  \Phi_{n+1} (z) + \\
&& \arcsin \Big[ \frac{P_n (z)}{\sqrt{P^2_n (z) + d^2_{12} N^2_n (z) }} \Big] \  \Big).
\label{pulse_area_gener_Phin}
\end{eqnarray}
When $P_n (z) \ne 0$, the obtained Eq.~\eqref{pulse_area_gener_Phin} cannot be generally integrated analytically. Therefore one has to rely on the numerical solution only.

The generalized area theorem Eq.~\eqref{pulse_area_gener} allows to find the pulse area and the medium parameters at the $(n+1)$-th round-trip, provided that these functions at the $n$-th round-trip are already known. However, since the equation Eq.~\eqref{pulse_area_gener_Phin} does not permit the analytical solution, it looks attracting to find out any reasonable simplifications to transform Eq.~\eqref{pulse_area_gener_Phin} into an analytically solvable form. The most natural way to do it is neglecting the spatial dependences of the medium parameters, i.e. setting $P_n (z) \approx \text{const}$, $N_n (z) \approx \text{const}$. With such an assumption Eq.~\eqref{pulse_area_gener_Phin} can be indeed easily solved to yield:
\begin{eqnarray}
\nonumber
\tan \Big(\frac{\Phi_{n+1} (z) + \tilde \Phi}{2} \Big)  &=& \tan \Big(\frac{\Phi_{n+1} (0) + \tilde \Phi}{2} \Big) \cdot \\
\nonumber
&&  e^{ \alpha z \sqrt{N^2_n + \frac{P^2_n }{d^2_{12}}} },  \\
\nonumber
\sin \tilde \Phi &=&  \frac{P_n }{\sqrt{P^2_n  + d^2_{12} N^2_n }}, \\
\nonumber
0  &\le&  z  \le  L_{\text{med}},  \\
\Phi_{n+1} (0)  &=&  \Phi_{n} (L_{\text{cav}}),
\label{pulse_area_gener_solution_simpl}
\end{eqnarray}
with the medium length $L_{\text{med}}$.

As can be seen, the analysis of the pulse propagation in a ring cavity  with Eq.~\eqref{pulse_area_gener_solution_simpl} reduces to a simple mapping connecting the pulse area and the medium parameters at the sequential round-trips. In the next Section we intend to examine the solutions of the generalized area theorem for the case provided by Eq.~\eqref{pulse_area_gener_solution_simpl} for a specific example of a single-section laser cavity with an amplifying medium inside.

\section{Single-section ring-cavity laser}

\begin{figure}[tb]
\centering
\includegraphics[width=0.95\linewidth]{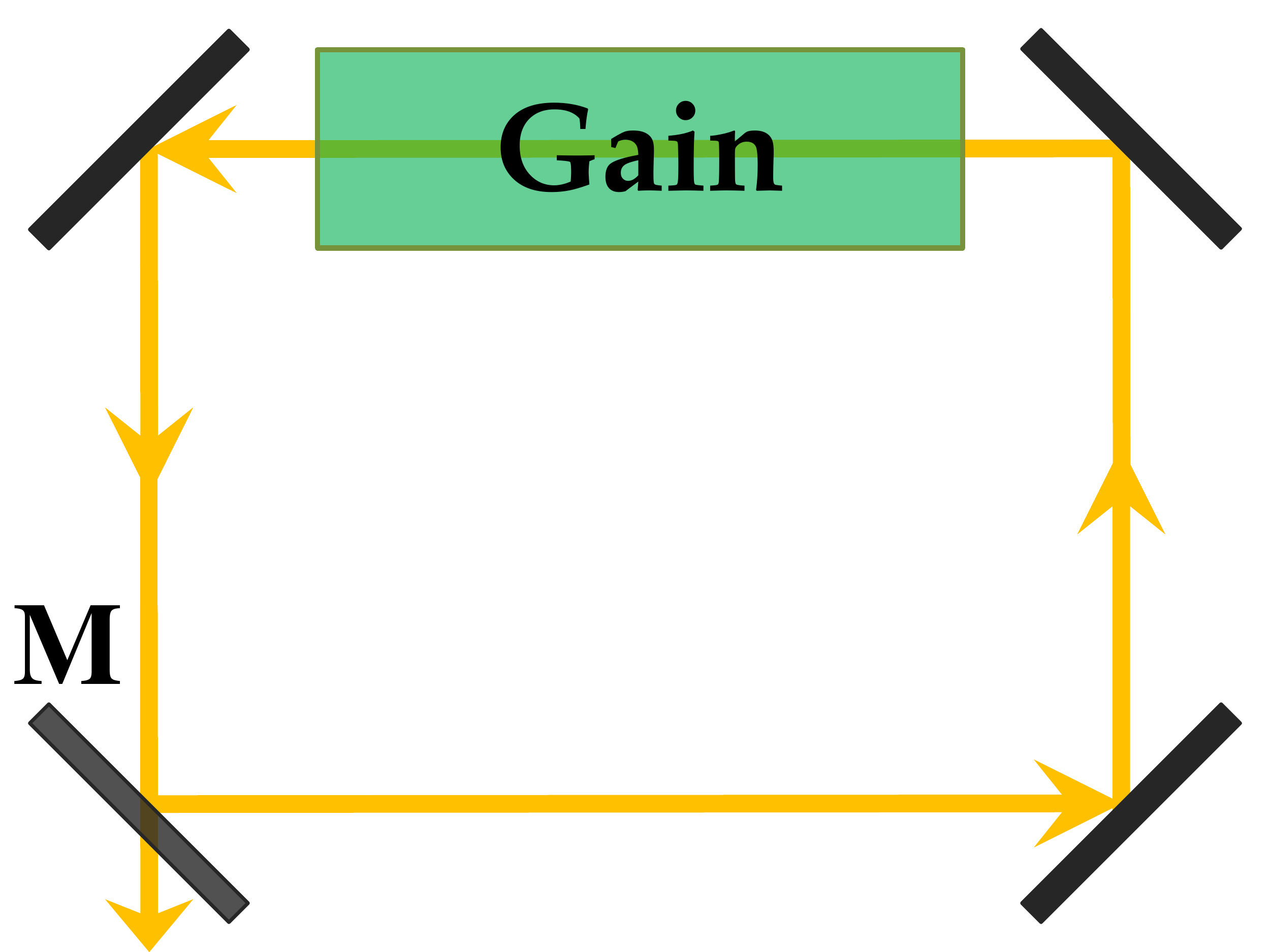}
\caption{The scheme of the considered single-section ring-cavity laser with the output mirror $M$; all other mirrors are assumed fully reflecting.}
\label{fig1}
\end{figure}

We start with a single-section ring cavity with the unidirectional field propagation, as sketched in Fig.~\ref{fig1}. It is assumed that the laser output is provided through one of the cavity mirrors with the amplitude reflection coefficient $r = r(\omega_{12})$ at the pulse carrier frequency $\omega_{12}$, while all other mirrors are fully reflecting. As before, the unidirectional lasing in the cavity is supposed, which can be achieved by placing an additional component inside the cavity.

%%%%%%%%%%%%%%%%%%%%%%%%%%%%%%%%%%%%%%%%%%%%%%%%%%%%%%%%%%%%
%%%%%%%%%%%%%%%%%%%%%%%%%%%%%%%%%%%%%%%%%%%%%%%%%%%%%%%%%%%%%

The setup shown in Fig.~\ref{fig1} can be exploited in several applications. Firstly, as it was numerically shown in \cite{Arkhipov_SciRep_2021}, the stable self-starting coherent mode-locking based on the $\pi$-pulse formation can arise in such a single-section laser with the achieved pulse duration of several hundreds of fs. However, the analytical study in \cite{Arkhipov_SciRep_2021} was solely focused on the case of a very long round-trip time, namely $T_1, \ T_2  \ll  T_{\text{rt}}$. At the same time taking a shorter cavity with $T_1, \ T_2  \sim  T_{\text{rt}}$ should lead both to much shorter mode-locked pulses and to much larger pulse repetition rate. Still the analysis for such cavity parameters can only be performed with the generalized area theorem derived in the previous section Eq.~\eqref{pulse_area_gener}.

Secondly, the laser cavity like in Fig.~\ref{fig1} can be naturally used for the efficient pulse compression. It is well known that the coherent pulse propagation in an amplifying medium results in the pulse amplification accompanied by the temporal compression \cite{kryukov1970propagation}. 
The cavity-based configuration then assures the multiple pulse passages through the gain medium and thus the sequential pulse compression.

It is worth noting that the pulse compression scheme needs the injection of an initial pulse into the cavity. The coherent mode-locking 
can also be started by the injection of an external seed pulse \cite{Kozlov_CML_2, Kozlov_CML_3}. However, the recent studies have also demonstrated that the coherent mode-locking can be self-starting, including the single-section geometry \cite{Arkhipov_OptCommun, Arkhipov_OL_2016, Arkhipov_SciRep_2021, Arkhipov_PRA_2022, Outafat, Pakhomov_PRA_2023}.
In the latter case the produced mode-locked pulse evolves from a weak initial noise in the
cavity. That is why in the following treatment, when talking about the pulse circulation inside a ring cavity, we will keep in mind that either an external seed pulse was initially injected into the cavity or some isolated field burst has randomly arisen against the weak background field.

For the setup in Fig.~\ref{fig1} the initial pulse area at the $(n+1)$-th round-trip is expressed as:
\begin{equation}
\Phi_{n+1} (0)  =  r \Phi_{n} (L_g),
\label{F_n+1_F_n}
\end{equation}
where $L_g$ is the length of the gain medium. 
Here we have used the result strictly proven in \cite{Arkhipov_SciRep_2021}, that the pulse areas Eq.~\eqref{eqPhi} of the incident and reflected ultra-short pulses upon the reflection from a mirror are simply related through the factor of the amplitude reflection coefficient $r$.

Below we will consider separately several cases, depending on the relative values of the relaxation times $T_1, T_2$ of the resonant medium and the cavity round-trip time $T_{\text{rt}}$.

A) $T_1, \ T_2  \ll  T_{\text{rt}}$.

This is the simplest possible situation, when the medium fully relaxes to its equilibrium state before the next pulse passage. Therefore the propagating pulse interacts at each round-trip with exactly the same medium, while the medium state right after the previous round-trip plays no role anymore.

Let us denote the pulse area just after the gain section after $n$ round-trips inside the cavity as $\Phi_n = \Phi_n (L_g)$. Now using Eq.~\eqref{pulse_area_gener_classic} and Eq.~\eqref{F_n+1_F_n} we get  the following explicit expression for the value $\Phi_{n+1}$:
\begin{equation}
\tan \Big( \frac{\Phi_{n+1}}{2} \Big)=\tan \Big(\frac{r \Phi_{n}}{2} \Big) \cdot e^{ \alpha N_{0, g} L_g}.
\label{F_n+1}
\end{equation}

The steady-state operation regime implies that:
\begin{equation}
\nonumber
\Phi_{n+1} = \Phi_{n} = \Phi^*,
\label{steady_state_def}
\end{equation}
so that we finally obtain the following equation for $\Phi^*$:
\begin{equation}
\Phi^* = 2 \arctan \Big[ e^{ \alpha N_{0, g} L_g} \cdot \tan \Big(\frac{r \Phi^*}{2} \Big) \Big]
\label{steady_state_eq}
\end{equation}

The solutions of Eq.~\eqref{steady_state_eq} and their stability were analysed in Ref.~\cite{Arkhipov_SciRep_2021}.
The performed analysis has shown that
Eq.~\eqref{steady_state_eq} always possesses the trivial solution:
\begin{equation}
\Phi^* = 0,
\label{trivial}
\end{equation} 
which is stable if $r e^{ \alpha N_{0, g} L_g}  <  1$, and the non-trivial steady state $\Phi^* \in (0; \pi)$, which exists if $ r e^{ \alpha N_{0, g} L_g}  >  1$ and is stable across the whole parameter range of its existence.

B) $T_2  \ll  T_{\text{rt}}  \sim  T_1$.

The induced medium polarization now completely vanishes over the round-trip. At the same time, the population inversion does not relax, so that its relaxation has to be properly considered.

As in the previous case, we denote the pulse area just after the gain section after $n$ round-trips inside the cavity as $\Phi_n$. Besides that, we introduce the population inversion in the gain section just before the $(n + 1)$-th passage of the pulse as $N_n$. For simplicity we start with the case of the constant population inversion over the whole gain section. This condition requires that the pulse area undergoes just slight changes during a single round-trip, i.e.:
\begin{equation}
\Big| \Phi_{n+1} - \Phi_n \Big|   \ll   \Big| \Phi_n \Big|.
\label{looped}
\end{equation}

The solution of the area theorem Eq.~\eqref{pulse_area_gener_reduced_solution} together with Eq.~\eqref{F_n+1_F_n} then gives the following expression connecting $\Phi_{n+1}$ and $\Phi_n$:
\begin{equation}
\Phi_{n+1} = 2 \arctan \Big[ e^{\alpha L_g N_n} \cdot \tan \Big(\frac{r \Phi_n}{2} \Big) \Big].
\label{Phi_n+1_1S_case2}
\end{equation}

As can be seen from Eq.~\eqref{Phi_n+1_1S_case2} the condition Eq.~\eqref{looped} implies that:
\begin{eqnarray}
\nonumber
\alpha L_g N_n  &\ll&  1, \\
1 - r  &\ll&  1.
\label{looped_2}
\end{eqnarray}

The expression for $N_{n+1}$ is provided by the second line in Eq.~\eqref{pulse_area_gener_reduced}, where the value $\Phi_{n+1} (z)$ in the right-hand side varies in between $r \Phi_n$ and the one given by Eq.~\eqref{Phi_n+1_1S_case2}. Using the inequalities Eqs.~\eqref{looped}, \eqref{looped_2}, we can reasonably take the average value of  $\Phi_{n+1} (z)$ equal to $\Phi_{n}$. As the result, Eq.~\eqref{pulse_area_gener_reduced} 
 yields for $N_{n+1}$:
\begin{eqnarray}
\nonumber
N_{n+1} = N_n \cos \Phi_n  \  e^{-T_{\text{rt}} / T_1 }  +  N_{0, g}  \Big( 1 - e^{-T_{\text{rt}} / T_1 } \Big). \\
\label{N_n+1_1S_case2}
\end{eqnarray}

Now we end up with the coupled equations Eqs.~\eqref{Phi_n+1_1S_case2}, ~\eqref{N_n+1_1S_case2}, which describe the evolution of the pulse and the active medium over the round-trips in the ring cavity.

Let us analyse the steady states of the obtained system Eqs.~\eqref{Phi_n+1_1S_case2}, ~\eqref{N_n+1_1S_case2}. One can easily see the trivial solution:
\begin{equation}
\Phi_n = \Phi^* = 0, \ \ \ \ N_n = N^* = N_{0, g}.
\label{trivial_1S_case2}
\end{equation}

We proceed now to check the stability of this trivial solution. We denote the functions in the right-hand side of Eqs.~\eqref{Phi_n+1_1S_case2}, ~\eqref{N_n+1_1S_case2} as $f_1(\Phi, N)$ and $f_2(\Phi, N)$ respectively. Then the stability of the steady state Eq.~\eqref{trivial_1S_case2} requires that:
\begin{equation}
| \lambda_i | <  1 \ \  \text{for all} \ \ \lambda_i  \in \text{eig} \ \textbf{J} ( f_1, f_2 ) \Big|_{\Phi = \Phi^*, \ N = N^*},
\label{stability_1S_case2}
\end{equation} 
where $\textbf{J}$ is the Jacobian matrix:
\begin{equation}
\textbf{J} ( f_1, f_2 ) =
\begin{vmatrix} 
\frac{\partial f_1}{\partial \Phi}  &  & \frac{\partial f_1}{\partial N}  \\
   &   &   \\
\frac{\partial f_2}{\partial \Phi}  &  & \frac{\partial f_2}{\partial N}
\end{vmatrix},
\label{Jacobian}
\end{equation} 
and $\lambda_i$ are the eigenvalues of this Jacobian matrix.

The condition Eq.~\eqref{stability_1S_case2} then reduces to:
\begin{equation}
\begin{cases}
\lambda_1 = r \ e^{\alpha L_g N_{0, g}}  <  1,  \\ 
\lambda_2 = e^{-T_{\text{rt}} / T_1 }  <  1.
\end{cases}
\label{stability_1S_case2_trivial}
\end{equation} 
As the second inequality in Eq.~\eqref{stability_1S_case2_trivial} is obeyed by default, only the first line Eq.~\eqref{stability_1S_case2_trivial} states the actual stability criteria for the trivial steady state. Note that this condition coincides with the one from the case (A).

Let us look for other non-trivial steady states of Eqs.~\eqref{Phi_n+1_1S_case2}, ~\eqref{N_n+1_1S_case2}. 
 i.e. the solutions in the form:
\begin{equation}
\Phi_n = \Phi^* \ne 0, \ \ \ \ N_n = N^* < N_{0, g}.
\label{nontrivial_1S_case2}
\end{equation}
Inspired by the findings for the case (A) we expect to get a steady state with the formation of almost $\pi$-pulses, i.e. $\Phi^* \approx \pi$. Inserting this approximate value into Eq.~\eqref{N_n+1_1S_case2} and taking $N_{n+1} = N_n = N^*$ one gets:
\begin{equation}
N^* = N_{0, g} \ \frac{1 - e^{-T_{\text{rt}} / T_1 } }{ 1 + e^{-T_{\text{rt}} / T_1 } }.
\label{nontrivial_N_1S_case2}
\end{equation}
According to Eq.~\eqref{nontrivial_N_1S_case2}, the faster is the inversion relaxation, the greater gain experiences the propagating pulse at each round-trip. Specifically, in the limit $T_{\text{rt}} \gg T_1$, i.e. the population inversion fully relaxes over a single round-trip, we arrive to the results of the case (A) with $N^* \to N_{0, g}$.

The stability of the non-trivial steady-state can be again checked using the criteria Eq.~\eqref{stability_1S_case2}. For the eigenvalues of the Jacobian matrix we find:
\begin{eqnarray}
\nonumber
 \lambda_1 &=& \frac{r \ e^{\alpha L_g N^*} }{  \cos^2 \Big( \frac{r \Phi^*}{2} \Big)  + 
 e^{2 \alpha L_g N^*} \ \sin^2 \Big( \frac{r \Phi^*}{2} \Big) }, \\
 \lambda_2  &=&  -e^{-T_{\text{rt}} / T_1 },
\label{stability_1S_case2_nontrivial_eigs}
\end{eqnarray} 
so that Eq.~\eqref{stability_1S_case2} gives:
\begin{eqnarray}
\begin{cases}
 r \ e^{\alpha L_g N^*}  <  \cos^2 \Big( \frac{r \Phi^*}{2} \Big)  + 
 e^{2 \alpha L_g N^*} \ \sin^2 \Big( \frac{r \Phi^*}{2} \Big), \\
 e^{-T_{\text{rt}} / T_1 }  <  1.
 \end{cases}
\label{stability_1S_case2_nontrivial}
\end{eqnarray} 
The second condition is again satisfied by default, while for the first one taking $r \approx 1, \ \Phi^* \approx \pi$ we obtain:
\begin{eqnarray}
\nonumber
 r  <  e^{\alpha L_g N^*},
\label{stability_1S_case2_nontrivial_2}
\end{eqnarray} 
which also always holds as $r<1$ and the exponent in the right-hand side is greater than 1, meaning that the non-trivial steady state Eq.~\eqref{nontrivial_N_1S_case2} is always stable.

The existence range of such non-trivial steady state can be analysed using the behaviour of the function in the right side of Eq.~\eqref{Phi_n+1_1S_case2}. 
Therefore we find that this non-trivial steady state exists when:
\begin{eqnarray}
 r \ e^{\alpha L_g N^*} > 1.
\label{existence_1S_case2_nontrivial_2}
\end{eqnarray} 
The comparison of the stability condition for the trivial solution provided by Eq.~\eqref{stability_1S_case2_trivial} and the existence condition of the non-trivial solution Eq.~\eqref{existence_1S_case2_nontrivial_2} yields, that because $N^* < N_{0, g}$ always, there is a range of the values of $N_{0, g}$, namely:
\begin{eqnarray}
\nonumber
\frac{1}{\alpha L_g} \log \Big( \frac{1}{r} \Big)  <  N_{0, g}   <   \frac{1}{\alpha L_g} \frac{ 1 + e^{-T_{\text{rt}} / T_1 }}{1 - e^{-T_{\text{rt}} / T_1 } } \ \log \Big( \frac{1}{r} \Big) , \\
\label{no_stability_range}
\end{eqnarray}
where the trivial state already becomes unstable, but the non-trivial steady state with $\Phi^* \approx \pi$ does not yet exist.
For example, for $T_{\text{rt}} = 0.1 T_1$ the right-hand side in Eq.~\eqref{no_stability_range} exceeds the left-hand one by the factor of 20.02, for $T_{\text{rt}} = T_1$ by the factor 2.16, for $T_{\text{rt}} = 0.5 T_1$ by the factor 4.08. In this range one can expect the formation of the non-trivial steady states with intermediate values of the pulse area $\Phi^* \in (0; \pi)$.

%%%%%%%%%%%%%%%%%%%%%%%%%%%%%%%%%%%%%%%%%%%%%%%%%%%%%%%%%%%%
\begin{figure}[t]
\centering
\includegraphics[width=1\linewidth]{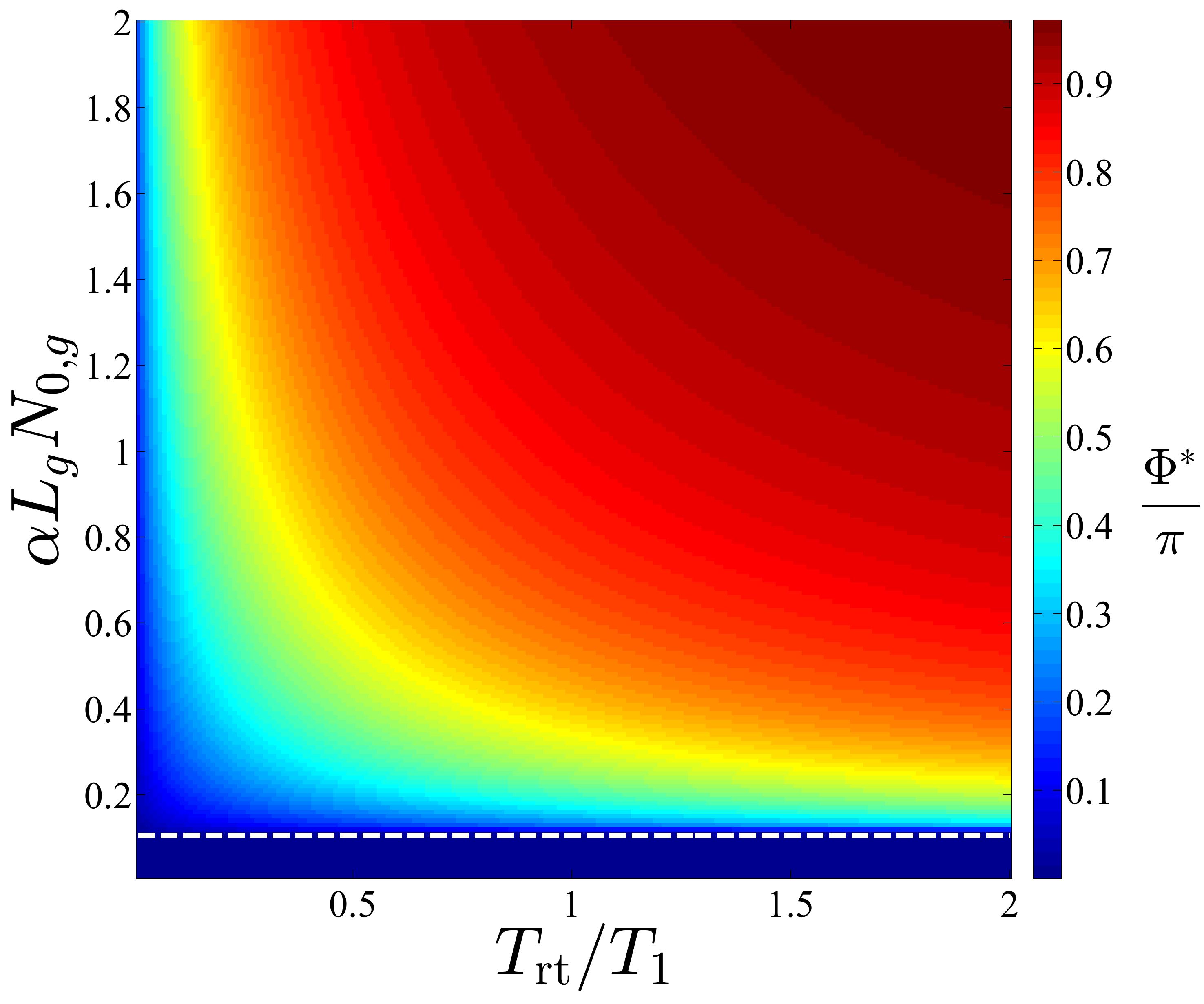}
\caption{The steady-state solution of Eqs.~\eqref{Phi_n+1_1S_case2}, \eqref{N_n+1_1S_case2} for the pulse area $\Phi^*$ vs. the parameters $T_{\text{rt}} / T_1$ and $\alpha L_g N_{0, g}$ for $r = 0.9$; the white dashed line depicts the stability threshold of the trivial steady state according to the first line in Eq.~\eqref{stability_1S_case2_trivial}.}
\label{fig2}
\end{figure}
%%%%%%%%%%%%%%%%%%%%%%%%%%%%%%%%%%%%%%%%%%%%%%%%%%%%%%%%%%%%%

%%%%%%%%%%%%%%%%%%%%%%%%%%%%%%%%%%%%%%%%%%%%%%%%%%%%%%%%%%%%
\begin{figure}[t]
\centering
\includegraphics[width=1\linewidth]{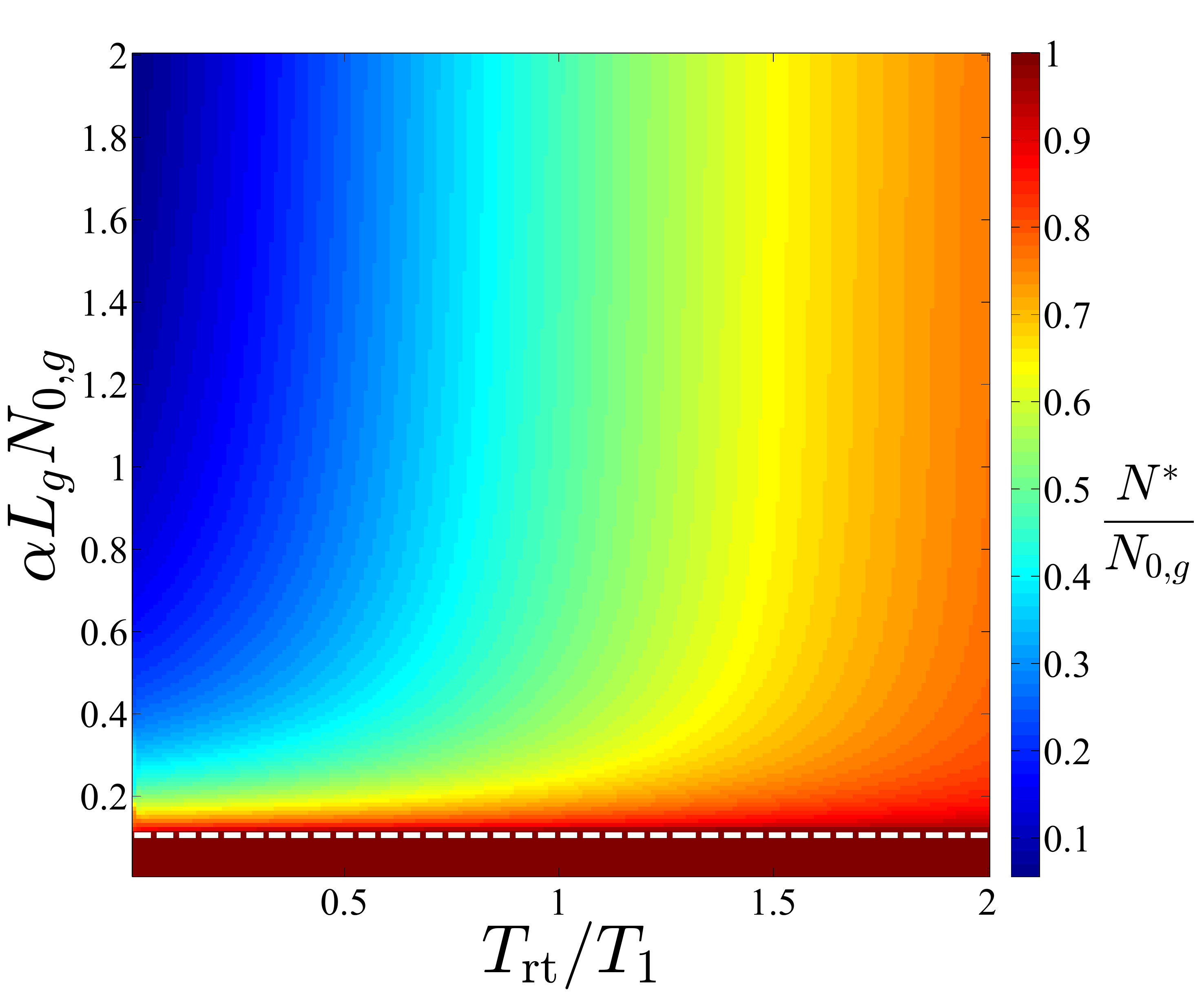}
\caption{The steady-state solution of Eqs.~\eqref{Phi_n+1_1S_case2}, \eqref{N_n+1_1S_case2} for the population inversion $N^*$  vs. the parameters $T_{\text{rt}} / T_1$ and $\alpha L_g N_{0, g}$ for $r = 0.9$; the white dashed line depicts the stability threshold of the trivial steady state according to the first line in Eq.~\eqref{stability_1S_case2_trivial}.}
\label{fig3}
\end{figure}
%%%%%%%%%%%%%%%%%%%%%%%%%%%%%%%%%%%%%%%%%%%%%%%%%%%%%%%

We have performed the numerical simulations of the mapping 
Eqs.~\eqref{Phi_n+1_1S_case2}, ~\eqref{N_n+1_1S_case2}, which showed that the system always rapidly reaches either the trivial or the non-trivial steady state.  Exemplary diagrams for the resulting steady-state solutions are plotted in Fig.~\ref{fig2} for the pulse area $\Phi^*$ and in Fig.~\ref{fig3} for the population inversion $N^*$. The reflection coefficient was fixed to $r = 0.9$, while the parameters $T_{\text{rt}} / T_1$ and $\alpha L_g N_{0, g}$  were varied. One can see that as the ratio $T_{\text{rt}} / T_1$ grows, the steady values go to $\Phi^* \to \pi$,  $N^* \to N_{0, g}$, what simply corresponds to the non-trivial solution of the case (A) with fast inversion relaxation.

The condition Eqs.~\eqref{looped}, ~\eqref{looped_2} is generally not satisfied over the first several iterations, if the initial conditions for $\Phi_0, \ N_0$ are taken quite far away from their steady values from Figs.~\ref{fig2}-\ref{fig3}. However, after a few iterations the system comes close enough to the steady states and for the following iterations the condition 
Eq.~\eqref{looped} is surely obeyed. This finding justifies the reduction of the spatially-distributed model from Section III to the simple mapping given by Eqs.~\eqref{Phi_n+1_1S_case2}, \eqref{N_n+1_1S_case2}, which is much more convenient both for the simulations and for the analysis.

C) $ T_{\text{rt}}  \sim  T_2, T_1$.

In this case the induced medium polarization does not make it to fully relax over the round-trip and the remaining polarization has to be taken into account.

As before, we assume for simplicity the constant population inversion and the medium polarization across the entire gain medium. 
The respective mapping for this case can be again obtained using the generalized area theorem Eq.~\eqref{pulse_area_gener} and Eq.~\eqref{pulse_area_gener_solution_simpl}.
Taking again the average value of $\Phi_{n+1} (z)$ equal to $\Phi_{n}$ in the equations for the medium quantities $N_{n+1}$ and $P_{n+1}$, we obtain:
\begin{eqnarray}
\nonumber
\Phi_{n+1} &=& -\arcsin \Big( \frac{P_n}{\sqrt{P_n^2 + d_{12}^2 N_n^2}} \Big)  +  \\
\nonumber
&& 2 \arctan \Big[ e^{\alpha L_g \sqrt{N_n^2 + P_n^2 / d_{12}^2 }} \cdot \\
\nonumber
&& \tan \Big( \frac{r}{2} \Phi_n + \frac{1}{2} \arcsin \Big( \frac{P_n}{\sqrt{P_n^2 + d_{12}^2 N_n^2}} \Big) \Big) \Big], \\
\nonumber
N_{n+1} &=& \Big( N_n \cos \Phi_n - \frac{1}{d_{12}} P_n \sin \Phi_n \Big)   e^{-T_{\text{rt}} / T_1 }  +  \\
\nonumber
&& N_{0, g}  \Big( 1 - e^{-T_{\text{rt}} / T_1 } \Big), \\
P_{n+1} &=& \Big( P_n \cos \Phi_n + d_{12} N_n \sin \Phi_n \Big)   e^{-T_{\text{rt}} / T_2 },
\label{Mapping_1S_case3}
\end{eqnarray}
where we have introduced the induced medium polarization after $n$ full round-trips right before the circulating pulse enters the gain section at the $(n+1)$-th iteration as $P_n$.

Similar to the previous cases the trivial steady state has the form:
\begin{equation}
\Phi_n = \Phi^* = 0, \ \ \ N_n = N^* = N_{0, g}, \ \ \ P_n = P^* = 0.
\label{trivial_1S_case3}
\end{equation}

For the eigenvalues of the Jacobian matrix Eq.~\eqref{Jacobian} of this trivial solution Eq.~\eqref{trivial_1S_case3}, one finds the explicit expression for one eigenvalue:
\begin{equation}
\nonumber
\lambda_1 = e^{-T_{\text{rt}} / T_1 } < 1,
\label{trivial_1S_case3_lamba1}
\end{equation}
while for two other eigenvalues $\lambda_2, \lambda_3$ the following characteristic equation arises:
\begin{eqnarray}
\nonumber
\lambda^2 + \lambda \ ( r e^{\alpha L_g N_{0, g}} + e^{-T_{\text{rt}} / T_2 } ) + \\
\nonumber
e^{-T_{\text{rt}} / T_2 } ( r e^{\alpha L_g N_{0, g}} + 1 - e^{\alpha L_g N_{0, g}}) = 0.
\label{trivial_1S_case3_lamba23}
\end{eqnarray}

%%%%%%%%%%%%%%%%%%%%%%%%%%%%%%%%%%%%%%%%%%%%%%%%%%%%%%%%%%%%
\begin{figure}[tb]
\centering
\includegraphics[width=1\linewidth]{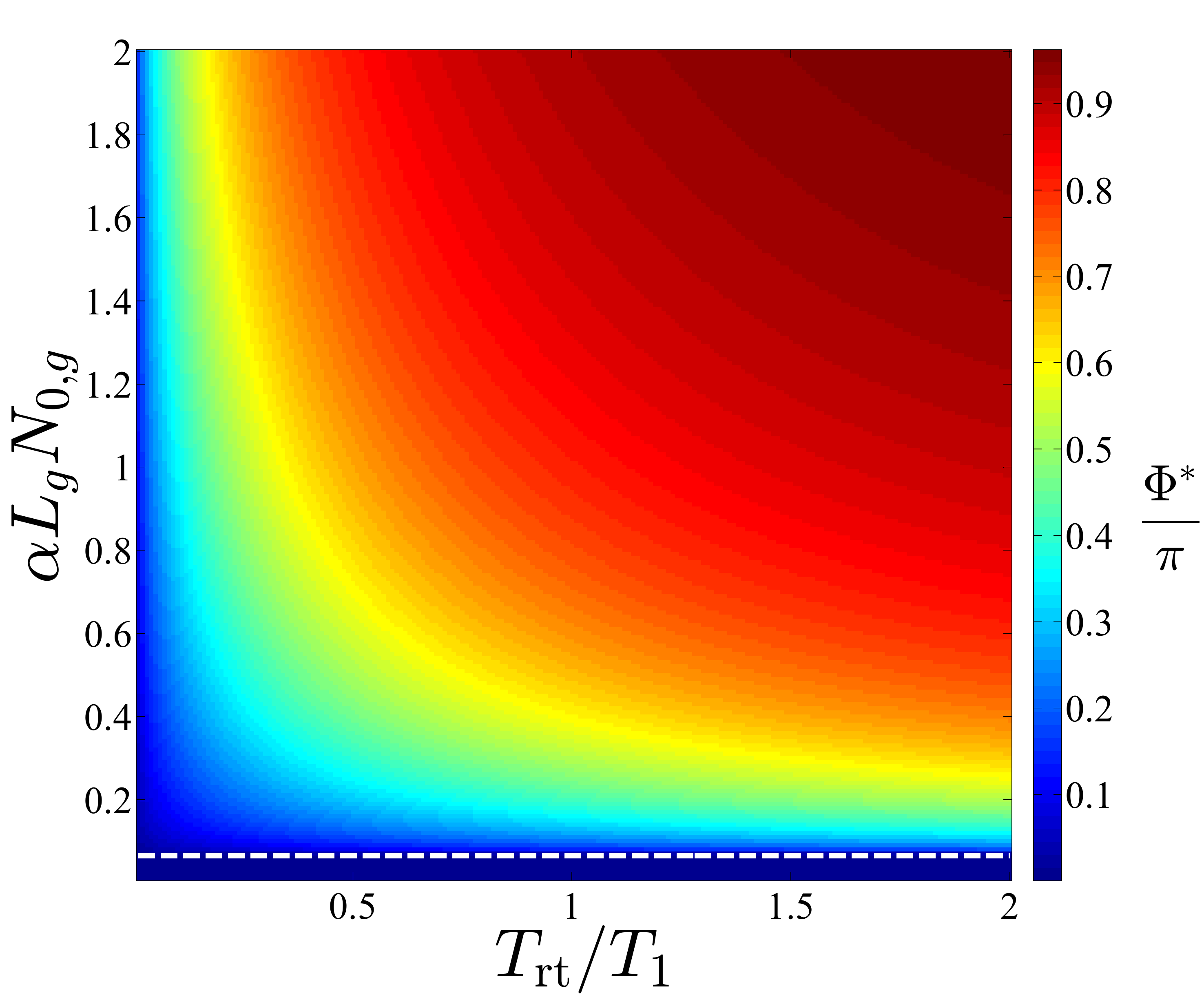}
\caption{The steady-state solution of Eq.~\eqref{Mapping_1S_case3} for the pulse area $\Phi^*$ vs. the parameters $T_{\text{rt}} / T_1$ and $\alpha L_g N_{0, g}$ for $r = 0.9$, $T_{\text{rt}} / T_2 = 1$; the white dashed line depicts the stability threshold of the trivial steady state according to Eq.~\eqref{trivial_1S_case3_lamba23_2}.}
\label{fig4}
\end{figure}
%%%%%%%%%%%%%%%%%%%%%%%%%%%%%%%%%%%%%%%%%%%%%%%%%%%%%%%%%%%%%

%%%%%%%%%%%%%%%%%%%%%%%%%%%%%%%%%%%%%%%%%%%%%%%%%%%%%%%%%%%%
\begin{figure}[tb]
\centering
\includegraphics[width=1\linewidth]{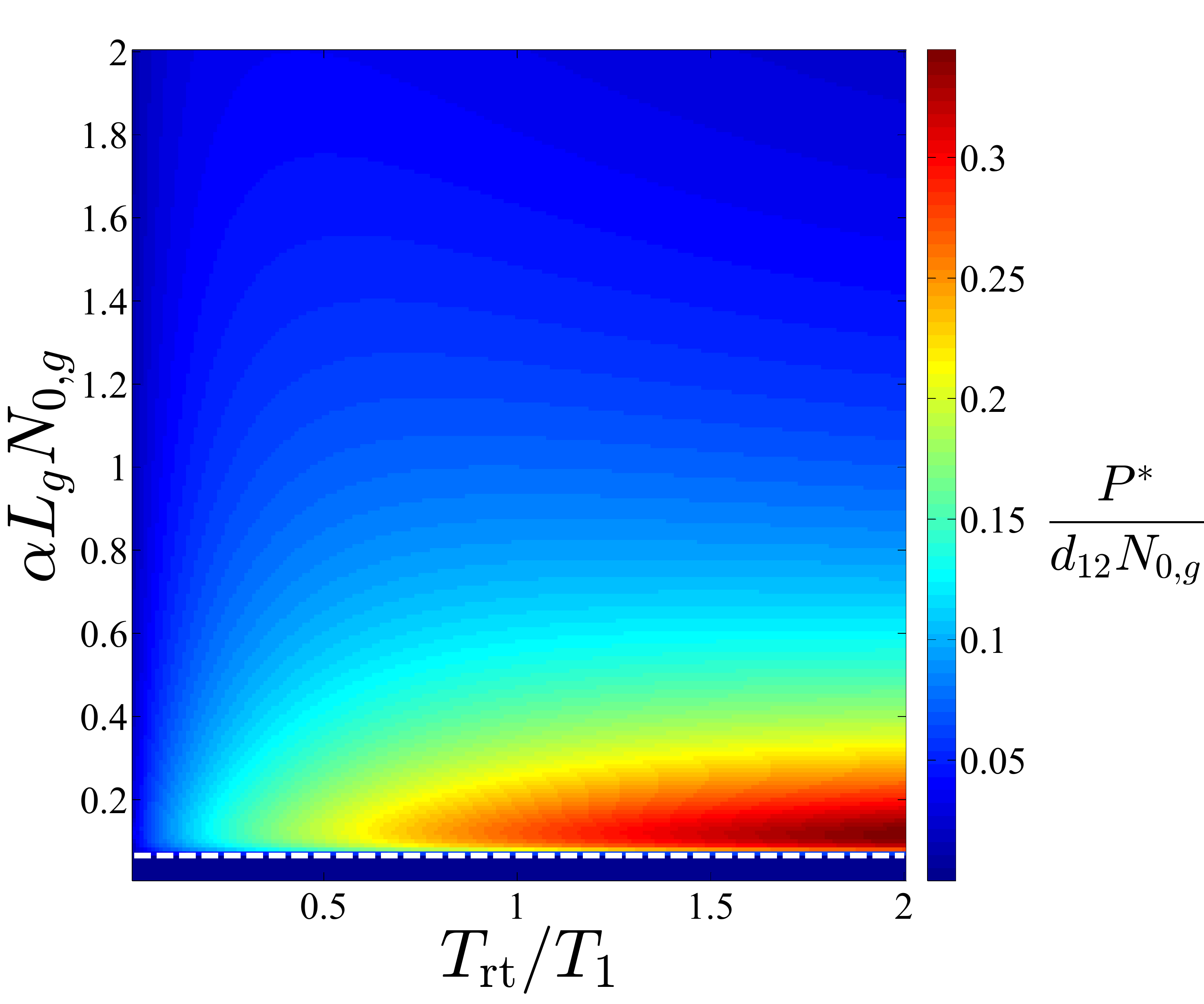}
\caption{The steady-state solution of Eq.~\eqref{Mapping_1S_case3} for the induced medium polarization $P^*$ vs. the parameters $T_{\text{rt}} / T_1$ and $\alpha L_g N_{0, g}$ for $r = 0.9$, $T_{\text{rt}} / T_2 = 1$; the white dashed line depicts the stability threshold of the trivial steady state according to Eq.~\eqref{trivial_1S_case3_lamba23_2}.}
\label{fig5}
\end{figure}
%%%%%%%%%%%%%%%%%%%%%%%%%%%%%%%%%%%%%%%%%%%%%%%%%%%%%%%

The analysis of this characteristic equation yields that the absolute values of both roots are less than unity as long as:
\begin{equation}
| \lambda_{2, 3} | < 1  \ \ \ \text{if}  \ \ \  e^{\alpha L_g N_{0, g}} ( r + e^{-T_{\text{rt}} / T_2 } (1-r) ) < 1. \\
\label{trivial_1S_case3_lamba23_2}
\end{equation}
In the limit $T_{\text{rt}} / T_2 \to +\infty$ we arrive to the case (B) above and the stability condition Eq.~\eqref{trivial_1S_case3_lamba23_2} reduces to the respective stability criteria Eq.~\eqref{stability_1S_case2_trivial}. From the comparison of Eq.~\eqref{stability_1S_case2_trivial} and Eq.~\eqref{trivial_1S_case3_lamba23_2} one can see that the slow relaxation of the medium polarization leads to the decrease of the lasing threshold.

Besides the trivial solution, the non-trivial steady state can be also expected to exist above the threshold Eq.~\eqref{trivial_1S_case3_lamba23_2}. Following the findings for the case (B), well above the threshold we can obtain the approximate solution for the non-trivial steady state as:
\begin{equation}
\Phi^* \approx \pi, \ \ \ P^* = 0, \ \ \ N^* = N_{0, g} \ \frac{1 - e^{-T_{\text{rt}} / T_1 } }{ 1 + e^{-T_{\text{rt}} / T_1 } }.
\label{nontrivial_1S_case3}
\end{equation}

The performed numerical simulations with the mapping Eq.~\eqref{Mapping_1S_case3} show that the system always rapidly evolves towards either the trivial solution Eq.~\eqref{trivial_1S_case3}, if the parameters obey the condition Eq.~\eqref{trivial_1S_case3_lamba23_2}, or towards the non-trivial steady state otherwise.

Fig.~\ref{fig4} depicts a diagram for the respective steady state of the pulse area in the mapping Eq.~\eqref{Mapping_1S_case3}, where the parameters were fixed to $r = 0.9$, $T_{\text{rt}} / T_2 = 1$, while the values of $T_{\text{rt}} / T_1$ and $\alpha L_g N_{0, g}$ were varied. Again, similar to the case (B), we get the stable trivial zero steady state below the threshold and the stable non-trivial steady state above the threshold, which goes to the solution Eq.~\eqref{nontrivial_1S_case3} as the pumping rate $N_{0, g}$ increases.

The respective diagram for the induced medium polarization $P^*$ is shown in Fig.~\ref{fig5}. Here one can see that the stable value of the polarization goes to zero as the pumping parameter $N_{0, g}$ gets larger, thus evolving towards the solution Eq.~\eqref{nontrivial_1S_case3}. As the result, we get a non-monotonic dependence of the stable value $P^*$ on the varying parameters. Namely, the steady-state solution for $P^*$ equals zero both below the threshold Eq.~\eqref{trivial_1S_case3_lamba23_2} and well above this threshold. The largest value of $P^*$ is therefore achieved just beyond the threshold Eq.~\eqref{trivial_1S_case3_lamba23_2}, as can be seen in Fig.~\ref{fig5}.

\begin{figure}[tb]
\centering
\includegraphics[width=1\linewidth]{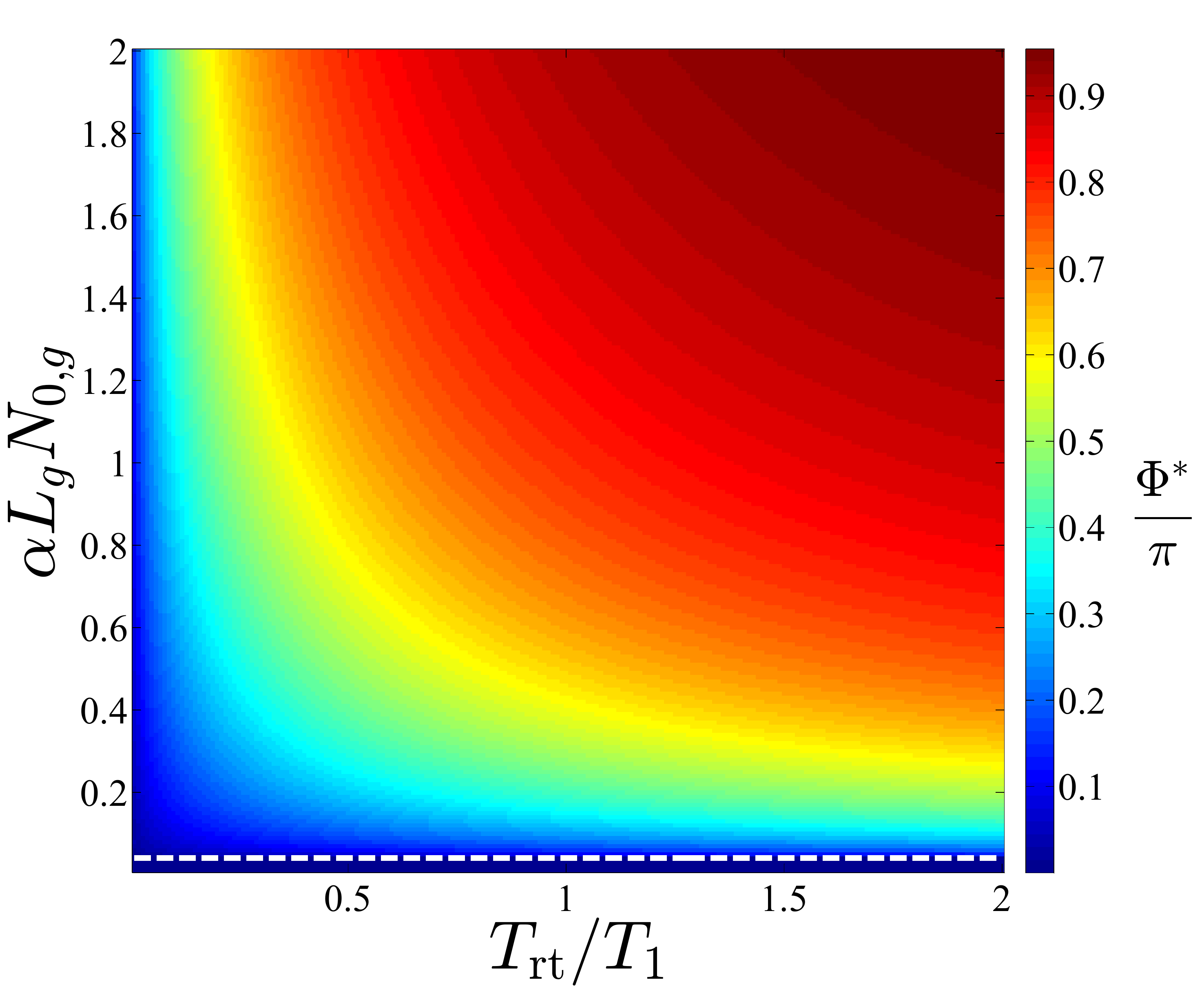}
\caption{The steady-state solutions of Eq.~\eqref{Mapping_1S_case3} for the pulse area $\Phi^*$ vs. the parameters $T_{\text{rt}} / T_1$ and $\alpha L_g N_{0, g}$ for $r = 0.9$, $T_{\text{rt}} / T_2 = 0.5$; the white dashed line depicts the stability threshold of the trivial steady state according to Eq.~\eqref{trivial_1S_case3_lamba23_2}.}
\label{fig6}
\end{figure}

It should be noted that the diagrams in Figs.~\ref{fig4}-\ref{fig5} do not exhibit significant deviations when changing the parameter $T_{\text{rt}} / T_2$. To illustrate this statement, we plot in Fig.~\ref{fig6} the similar diagram for the stable pulse area $\Phi^*$ like in Fig.~\ref{fig4}, but with the ratio $T_{\text{rt}} / T_2$ reduced to $T_{\text{rt}} / T_2 = 0.5$. The resulting behaviour of the steady-state solution stays largely unchanged, indicating the stability of the obtained solutions against the variations of the system parameters.

\section{Spatially-extended solutions}

In the previous section, specifically for cases (B) and (C), we made use of the approximations of the spatially-homogeneous functions $P_n (z) \approx \text{const}$, $N_n (z) \approx \text{const}$. Let us now consider in more details the role of the spatial extension in the dynamics of a single-section ring laser cavity and find out the validity limits for such approximations. Thereby we study the solution of the generalized pulse area theorem Eq.~\eqref{pulse_area_gener} taking into account the spatial coordinate $z$.

The case (A) from the previous section is trivial, since the inversion fully relaxes back to $N_{0, g}$ and the induced medium polarization fully vanishes, so that in the stable regime we end up with the spatially-homogeneous solutions:
\begin{equation}
\nonumber
N^* (z) = N_{0, g} = \text{const}, \ \ \ P^* (z) = 0 = \text{const}.
\label{solution_z_1S_case1}
\end{equation}

%%%%%%%%%%%%%%%%%%%%%%%%%%%%%%%%%%%%%%%%%%%%%%%%%%%%%%%%%%%%
%%%%%%%%%%%%%%%%%%%%%%%%%%%%%%%%%%%%%%%%%%%%%%%%%%%%%%%%%%%%%

That is why we are going to focus on the case (B), when the dynamics of the population inversion comes into play. Specifically, we are interested to find the solutions for $\Phi^* (z)$ and $N^* (z)$ in the stable regime of the pulse circulation inside the cavity. For this aim we insert into the pulse area theorem Eqs.~\eqref{pulse_area_gener_reduced}-\eqref{pulse_area_gener_reduced_solution}:
\begin{eqnarray}
\nonumber
\Phi_{n+1} (z) = \Phi_n (z) = \Phi^* (z), \\
N_{n+1} (z) = N_n (z) = N^* (z).
\label{solution_z_1S_case2}
\end{eqnarray}

Substituting the expressions Eq.~\eqref{solution_z_1S_case2} into Eqs.~\eqref{pulse_area_gener_reduced}-\eqref{pulse_area_gener_reduced_solution} we get the following integral equations for the unknown functions $N^* (z)$ and $\Phi^* (z)$:
\begin{eqnarray}
\nonumber
N^* (z) &=& N^* (z) \cos \Phi^* (z)  \  e^{-T_{\text{rt}} / T_1 }  +  \\
\nonumber
&& N_{0, g}  \Big( 1 - e^{-T_{\text{rt}} / T_1 } \Big), \\
\nonumber
\Phi^* (z) &=& 2 \arctan \Big[ e^{\alpha  \int_0^z N^* (z') dz'} \cdot \tan \Big(\frac{\Phi^* (0)}{2} \Big) \Big], \\
0 &\le& z \le L_g.
\label{solution_z_1S_case2_2}
\end{eqnarray}

Using the formula:
\begin{equation}
\nonumber
\cos \varphi = \frac{ 1 - \tan^2 \frac{\varphi}{2}}{1 + \tan^2 \frac{\varphi}{2}}, 
\label{tangent_x_2}
\end{equation}
we can reduce the equations Eq.~\eqref{solution_z_1S_case2_2} to:
\begin{eqnarray}
\nonumber
\cos \Phi^* (z)  &=&  e^{T_{\text{rt}} / T_1 } - \frac{N_{0, g}}{N^* (z)}  \Big( e^{T_{\text{rt}} / T_1 } - 1 \Big), \\
\nonumber
\cos \Phi^* (z)  &=&  \frac{ 1 - e^{ 2 \alpha \int_0^z N^* (z') dz'} \cdot \tan^2 \Big(\frac{\Phi^* (0)}{2} \Big) }{ 1 + e^{ 2 \alpha \int_0^z N^* (z') dz'} \cdot \tan^2 \Big(\frac{\Phi^* (0)}{2} \Big) }, \\
\nonumber
0 &\le& z \le L_g.
\label{solution_z_1S_case2_3}
\end{eqnarray}

Putting the right-hand sides of the above two expressions equal, we get:
\begin{eqnarray}
\nonumber
e^{ 2 \alpha \int_0^z N^* (z') dz'} \cdot \tan^2 \Big(\frac{\Phi^* (0)}{2} \Big) = \\
\nonumber
\frac{N^* (z) \Big( 1 - e^{T_{\text{rt}} / T_1 } \Big) + N_{0, g} \Big( e^{T_{\text{rt}} / T_1 } - 1 \Big) }{N^* (z) \Big( 1 + e^{T_{\text{rt}} / T_1 } \Big) - N_{0, g} \Big( e^{T_{\text{rt}} / T_1 } - 1 \Big)}, \\
0 \le z \le L_g.
\label{solution_z_1S_case2_4}
\end{eqnarray}

Now taking the log of both sides of the equation Eq.~\eqref{solution_z_1S_case2_4} and then differentiating over $z$, we find:
\begin{eqnarray}
\nonumber
&&  2 \alpha N^* (z)  =  \frac{d N^* (z)}{dz} \cdot \frac{2 N_{0, g} }{\Big( N^* (z) - N_{0, g} \Big) } \cdot \\
\nonumber
&& \frac{1}{\Big( N^* (z) \Big( 1 + e^{T_{\text{rt}} / T_1 } \Big) - N_{0, g} \Big( e^{T_{\text{rt}} / T_1 } - 1 \Big) \Big)}, \\
&&  0 \le z \le L_g.
\label{solution_z_1S_case2_5}
\end{eqnarray}

It is convenient to rewrite the obtained ordinary differential equation Eq.~\eqref{solution_z_1S_case2_5} in the form:
\begin{eqnarray}
\frac{d N^*}{N^* (N^* - N_{0, g}) (N^* - R)}  = \kappa dz,
\label{solution_z_1S_case2_6}
\end{eqnarray}
with the following parameters:
\begin{eqnarray}
\nonumber
R  &=&  N_{0, g} \ \frac{e^{T_{\text{rt}} / T_1 } - 1}{e^{T_{\text{rt}} / T_1 } + 1} > 0, \\
\nonumber
\kappa  &=&  \frac{\alpha}{N_{0, g}} \ \Big( 1 + e^{T_{\text{rt}} / T_1 } \Big).
\label{parameters_1S_case2}
\end{eqnarray}

Assuming that:
\begin{equation}
\nonumber
 R < N^* (z) < N_{0, g},
\label{solution_z_1S_case2_correctness}
\end{equation}
we can integrate Eq.~\eqref{solution_z_1S_case2_6} as follows:
\begin{eqnarray}
\nonumber
&&  \log \Big( \frac{N^*(z)}{N^*(0)} \Big)  +  \frac{R}{N_{0, g} - R} \log \Big( \frac{N_{0, g} - N^*(z)}{N_{0, g}  - N^*(0)} \Big)  -  \\
\nonumber
&&  \frac{N_{0, g}}{N_{0, g} - R} \log \Big( \frac{N^*(z) - R}{N^*(0) - R} \Big)   =   \kappa N_{0, g} R \ z, \\
&&  0 \le z \le L_g.
\label{solution_z_1S_case2_final}
\end{eqnarray}

The integration constant  $N^*(0)$ can be expressed from the first line in Eq.~\eqref{solution_z_1S_case2_2} through the value $\cos \Phi^* (0)$ as:
\begin{equation}
N^* (0)  =  N_{0, g} \ \frac{1 - e^{-T_{\text{rt}} / T_1 }}{1 - \cos \Phi^* (0)  \  e^{-T_{\text{rt}} / T_1 }  }.
\label{solution_z_integr_const}
\end{equation}

%%%%%%%%%%%%%%%%%%%%%%%%%%%%%%%%%%%%%%%%%%%%%%%%%%%%%%%%%%%%
\begin{figure}[tb]
\centering
\includegraphics[width=1\linewidth]{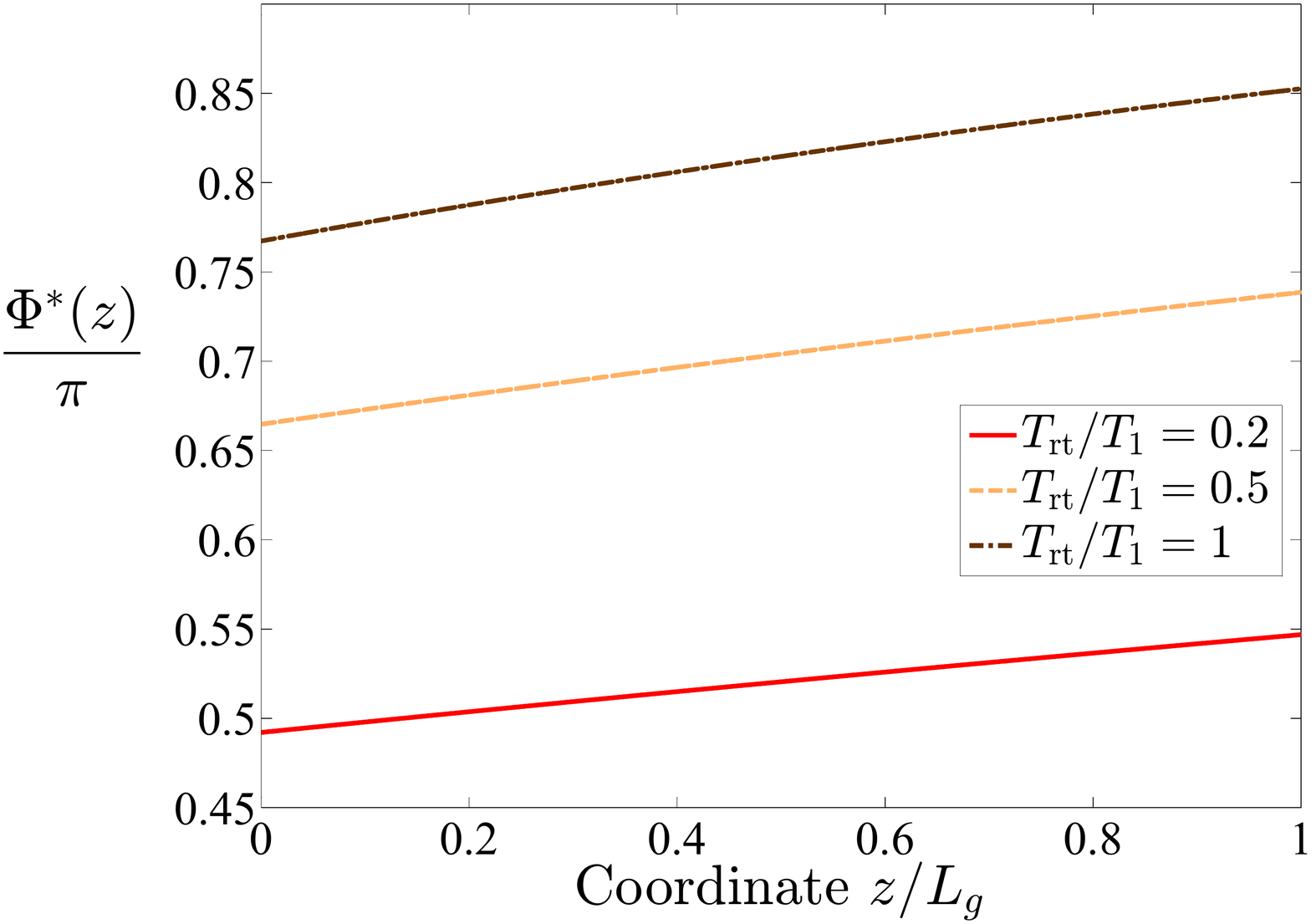}
\caption{The steady-state spatially-extended solutions of Eq.~\eqref{solution_z_1S_case2_2} for the pulse area $\Phi^* (z)$ for the parameter values $r = 0.9$, $\alpha L_g N_{0, g} = 1$  and several different values of the ratio $T_{\text{rt}} / T_1$.}
\label{fig7}
\end{figure}
%%%%%%%%%%%%%%%%%%%%%%%%%%%%%%%%%%%%%%%%%%%%%%%%%%%%%%%%%%%%%

Now using the relation between the values of the pulse area at the boundaries in the steady-operation regime:
\begin{equation}
\nonumber
\Phi^* (0) = r \Phi^* (L_g)
\label{Phi_Lg_0}
\end{equation}
we can rewrite the second line in Eq.~\eqref{solution_z_1S_case2_2} taking $z = L_g$:
\begin{eqnarray}
\nonumber
\frac{\Phi^* (0)}{r}  = 2 \arctan \Big[ e^{\alpha  \int_0^{L_g} N^* (z', \Phi^* (0)) dz'} \cdot \tan \Big(\frac{\Phi^* (0)}{2} \Big) \Big]. \\
\label{solution_z_integr_const_Phi}
\end{eqnarray}
Here we explicitly stated that the steady solution for 
$N^*(z)$ depends on the value $\Phi^* (0)$ through the expression for the integration constant 
 $N^*(0)$  Eq.~\eqref{solution_z_integr_const}.

Solving numerically Eq.~\eqref{solution_z_integr_const_Phi} for the value $\Phi^* (0)$, we eventually get the sought-for solutions for $N^* (z)$ from Eqs.~\eqref{solution_z_1S_case2_final}-\eqref{solution_z_integr_const} and the respective solution for $\Phi^* (z)$ from the second line in Eq.~\eqref{solution_z_1S_case2_2}.

\begin{figure}[t]
\centering
\includegraphics[width=1\linewidth]{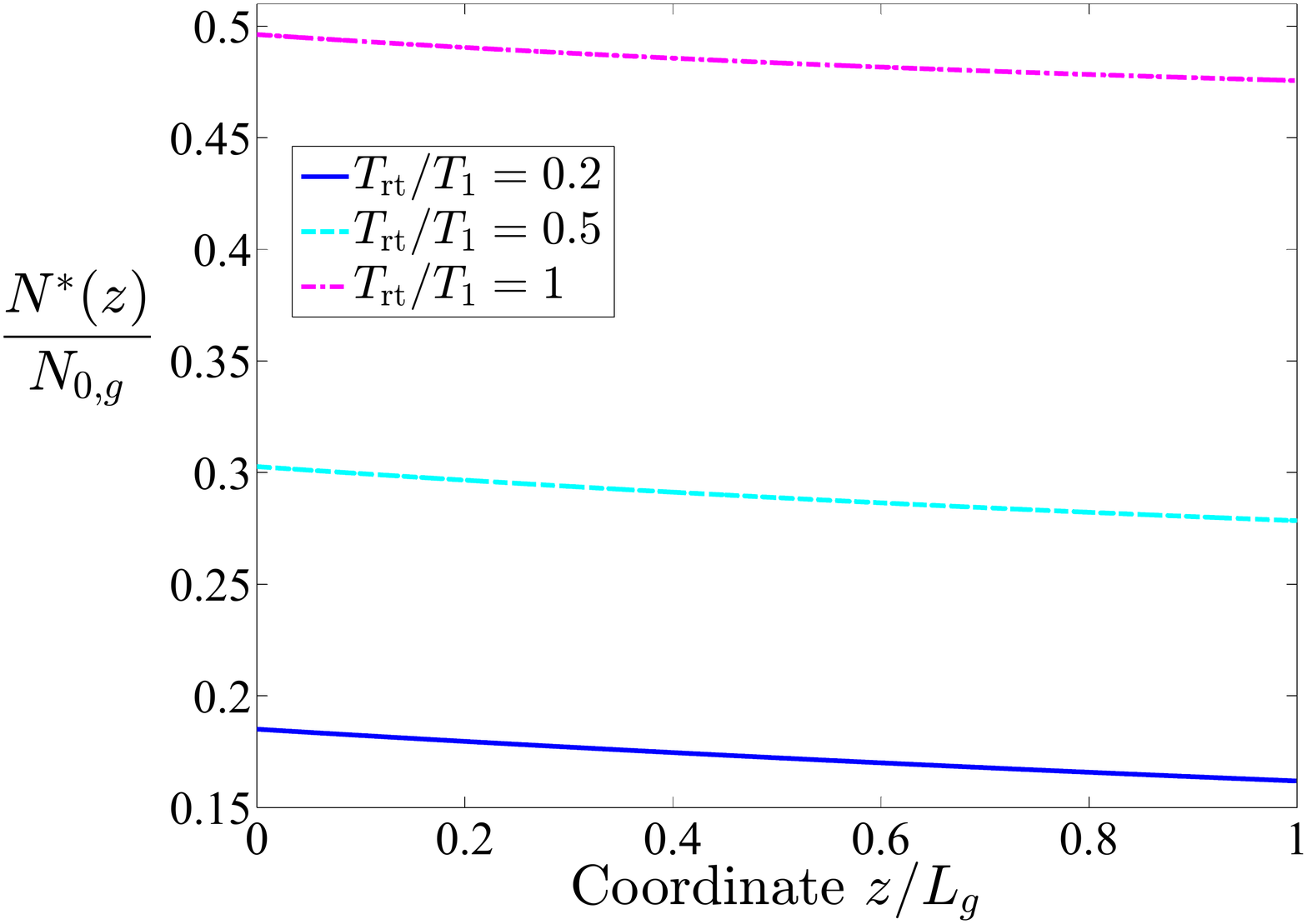}
\caption{The steady-state spatially-extended solutions of Eq.~\eqref{solution_z_1S_case2_2} for the population inversion $N^* (z)$ for the parameter values $r = 0.9$, $\alpha L_g N_{0, g} = 1$  and several different values of the ratio $T_{\text{rt}} / T_1$.}
\label{fig8}
\end{figure}

A few examples of the resulting steady-state solutions are plotted in Figs.~\ref{fig7}-\ref{fig8} for the functions $\Phi^* (z)$ and $N^* (z)$ respectively. In these figures the values of $r$ and $\alpha L_g N_{0, g}$ were fixed, while the ratio $T_{\text{rt}} / T_1$ was varied. The obtained values of both functions $\Phi^* (z)$ and $N^* (z)$ turn out to closely match the steady-state values of $\Phi^*$ and $N^*$ for the lumped model with the same parameters, as shown in  Figs.~\ref{fig2}-\ref{fig3}.

%%%%%%%%%%%%%%%%%%%%%%%%%%%%%%%%%%%%%%%%%%%%%%%%%%%%%%%%%%%%
%%%%%%%%%%%%%%%%%%%%%%%%%%%%%%%%%%%%%%%%%%%%%%%%%%%%%%%%%%%%%

As Fig.~\ref{fig8} shows, the steady-state spatially-varying solutions for $N^* (z)$ in a wide range of the laser parameters exhibit rather limited spatial inhomogeneity. Specifically, the relative variations of the functions $N^* (z)$ across the entire length of the medium would barely exceed 10 \%. This finding can thus serve as the justification of the approximations made in the Section IV, where spatially-homogeneous functions $P_n (z) \approx \text{const}$, $N_n (z) \approx \text{const}$ were assumed.

%\section{Two-section ring-cavity laser}

%We proceed now with a two-section ring-cavity laser configuration, with a gain and an absorber section placed separately inside the cavity, as sketched in Fig.~\ref{figX}. 

%%%%%%%%%%%%%%%%%%%%%%%%%%%%%%%%%%%%%%%%%%%%%%%%%%%%%%%%%%%%
%\begin{figure}[tb]
%\centering
%\includegraphics[width=0.95\linewidth]{FigureX.pdf}
%\caption{The scheme of the considered two-section ring-cavity laser with the output mirror $M$; all other mirrors are assumed fully reflecting.}
%\label{figX}
%\end{figure}
%%%%%%%%%%%%%%%%%%%%%%%%%%%%%%%%%%%%%%%%%%%%%%%%%%%%%%%%%%%%%

% We are thus interested to find other possible stable limit cycles of Eqs.~\eqref{Phi_n+1_1S_case2}-\eqref{N_n+1_1S_case2}. Let us start with the case of a 2-cycle. It means we are looking for the solutions of the form:
%\begin{eqnarray}
%\nonumber
%\Phi_{n+2}  &=& \Phi_n  =  \Phi^*, \\
%N_{n+2}  &=&  N_n  =  N^*,
%\label{2cycle_1S_case2}
%\end{eqnarray} 
%what gives the following system:
%\begin{eqnarray}
%\nonumber
%\Phi^*  &=&  f_1 \Big( f_1(\Phi^*, N^*), f_1(\Phi^*, N^*) \Big), \\
%N^*  &=&  f_2 \Big( f_1(\Phi^*, N^*), f_1(\Phi^*, N^*) \Big).
%\label{2cycle_1S_case2_2}
%\end{eqnarray} 

\section{Conclusion}

We have derived the generalized area theorem for the unidirectional pulse circulation in a ring laser cavity. This generalization allows to describe the feedback action of the residual medium excitation, i.e. the residual induced polarization and the population inversion, caused by the pulse at the previous round-trip in the cavity on the pulse propagation at the next round-trip. Such effects are essential, as long as the round-trip time of the cavity becomes comparable to the medium relaxation times, e.g. in compact lasers.

The obtained generalized area theorem includes the ordinary differential equation for the spatial evolution of the pulse area at each round-trip inside a ring cavity together with two explicit expressions for the spatially-dependent medium polarization and the population inversion. In the general case, when the dynamics of both the medium polarization and the population inversion has to be taken into account, the equation for the pulse area can not be analytically integrated. For this case we have considered the reduction of the derived area theorem to the lumped system, assuming the medium quantities roughly constant in space. As the result, the theoretical model turns into a simple mapping for the pulse area $\Phi_{n+1}$ and the medium quantities $N_{n+1}$ and $P_{n+1}$ expressed through their values at the previous round-trip.

The derived area theorem as well as its reduced lumped form were tested with an exemplary setup of a single-section ring-cavity laser with the gain section only, which could represent either a single-section coherently mode-locked laser or a multi-pass cavity-based pulse compressor. The performed analysis for different relations between the relaxation times $T_1$, $T_2$ and the round-trip time $T_{\text{rt}}$ has yielded that the system always rapidly evolves towards either the trivial steady state below a certain lasing threshold, or to  a non-trivial steady state above the threshold. It is also worth noting that the stable value of the pulse area tends to $\pi$ well above the threshold. Beside that, the lumped version of the area theorem was shown to yield satisfactory quantitative agreement with the spatially-extended model.

The considered above approach is quite similar to that used in standard passive mode-locking theories based on incoherent light-matter interactions \cite{haus1975theory, haus1975theory_2, haus2000mode, new1974pulse, kartner1996soliton, kurtner1998mode, paschotta2001passive, vladimirov2004new, vladimirov2004delay, vladimirov2005model}, %\cite{haus1975theory, haus1975theory_2, haus2000mode, new1974pulse, kartner1996soliton, kurtner1998mode, paschotta2001passive, vladimirov2004new, vladimirov2004delay,vladimirov2005model, arkhipov2012hybrid,arkhipov2015pulse, arkhipov2016semiconductor, vladimirov2021delay,vladimirov2022short}, 
where evolution of the field amplitude at each round-trip is expressed through its value at the previous round-trip. However, these theories are inapplicable for the correct description of passively mode-locked ultrafast laser systems, where coherent effects play crucial role. It was shown earlier that to obtain shorter pulses it is necessary to decrease the cavity length \cite{Arkhipov_SciRep_2021, Arkhipov_PRA_2022}. In such compact laser systems the pulse duration is already comparable to the medium coherence lifetime $T_2$ and the coherent effects are of significant importance. Thus, the area theorem approach introduced above can be useful for the theoretical description of mode-locking in ultrafast laser systems, where  previously developed mode-locking theories are not applicable anymore.

We expect that the presented generalized area theorem can be used for the efficient analytical description of the coherent mode-locking phenomena, the cavity-assisted pulse compressors based on the coherent pulse propagation or for the description of photon-echo effects in cavity-based  schemes, e.g. in view of quantum-memory applications.

\section*{Acknowledgements}
The authors acknowledge support from the Foundation for the Advancement of Theoretical Physics and Mathematics “BASIS”.

% Bibliography
\bibliography{UP_library}

\end{document}